\documentclass[a4paper, 11pt]{article}
\usepackage{amsmath,amssymb,amsthm}
\usepackage{dsfont}
\usepackage{graphicx}
\usepackage{subcaption}
\usepackage[british]{babel}
\usepackage[utf8]{inputenc}
\usepackage{xcolor}
\usepackage[colorlinks=true,linkcolor=blue,citecolor=red]{hyperref}
\usepackage{authblk}
\usepackage{blindtext}
\usepackage{epstopdf}
\usepackage{cite}
\usepackage{bm}
\usepackage{comment}
\usepackage{booktabs} 
\usepackage{algorithm}
\usepackage[shortlabels]{enumitem}
\usepackage{tikz}
\usepackage{hhline}
\usetikzlibrary{calc,trees,positioning,arrows,chains,shapes.geometric,%
  decorations.pathreplacing,decorations.pathmorphing,shapes,%
  matrix,shapes.symbols}
\tikzset{
>=stealth',
 punktchain/.style={
  rectangle, 
   fill=cyan!40,
  draw=black, very thick,
  text width=12em, 
  minimum height=2em, 
  text centered, 
  on chain},
 line/.style={draw, thick, <-},
 element/.style={
  tape,
  top color=white,
  bottom color=blue!50!black!60!,
  minimum width=8em,
  draw=blue!40!black!90, very thick,
  text width=10em, 
  minimum height=2.5em, 
  text centered, 
  on chain},
 every join/.style={->, thick,shorten >=1pt},
 decoration={brace},
 tuborg/.style={decorate},
 tubnode/.style={midway, right=2pt},
}

\newtheorem{remark}{Remark}

\usepackage{mathtools}

\newcommand{\R}{\mathbb{R}}

\newcommand{\E}{\mathbb{E}}

\newcommand{\be}{\begin{equation}}
\newcommand{\ee}{\end{equation}}
\newcommand{\z}{\mathbf{z}}

\renewcommand{\epsilon}{\varepsilon}
\newcommand{\rev}[1]{\textcolor{black}{#1}}

\newcommand{\bk}{\mathbf{k}}
\newcommand{\bx}{\mathbf{x}}
\newcommand{\eps}{\varepsilon}

\allowdisplaybreaks

\begin{document}
\title{Uncertainty quantification for charge transport in GNRs through particle Galerkin methods for the semiclassical Boltzmann equation}

\author[1]{Andrea Medaglia\thanks{\tt andrea.medaglia@maths.ox.ac.uk }}
\author[2]{Giovanni Nastasi\thanks{\tt giovanni.nastasi@unikore.it}}
\author[3]{Vittorio Romano\thanks{\tt romano@dmi.unict.it }}
\author[4]{Mattia Zanella\thanks{\tt mattia.zanella@unipv.it}}
\affil[1]{Mathematical Institute, University of Oxford, United Kingdom}
\affil[2]{Department of Engineering and Architecture, University of Enna ``Kore", Italy}
\affil[3]{Department of Mathematics and Computer Science, University of Catania, Italy}
\affil[4]{Department of Mathematics ``F. Casorati", University of Pavia, Italy}

\date{}

\maketitle

\abstract{
In this article, we investigate some issues related to the quantification of uncertainties associated with the electrical properties of graphene nanoribbons. The approach is suited to understand the effects of missing information linked to the difficulty of fixing some material parameters, such as the band gap, and the strength of the applied electric field. In particular, we focus on the extension of particle Galerkin methods for kinetic equations in the case of the semiclassical Boltzmann equation for charge transport in graphene nanoribbons with uncertainties.
To this end, we develop an efficient particle scheme which allows us to parallelize the computation  and then, after a suitable generalization of the scheme to the case of random inputs, we present a Galerkin reformulation of the particle dynamics, obtained by means of a generalized \rev{Polynomial Chaos} approach, which allows the reconstruction of the kinetic distribution. As a consequence, the proposed particle-based scheme preserves the physical properties and the positivity of the distribution function also in the presence of a complex scattering in the transport equation of electrons. The impact of the uncertainty of the band gap and applied field on the electrical current is analysed. 
}
\\[+.2cm]
\\[+.2cm]
\tableofcontents

\section{Introduction}
Low dimensional materials are investigated in electronics with the aim to reduce the dimensions of the future  electronic devices. One of the most prominent 2D materials is graphene for its very peculiar electronic properties. The absence of an energy gap limits the use of graphene in field effect transistors (FETs) because there exists a restricted current-off region. A possible way to overcome such a drawback is to consider narrow strips of graphene, called graphene nanoribbons (GNRs), see e.g. \cite{chen_etal}. In fact, the spatial confinement induces a band gap, even if the mobility reduces with respect to the large area graphene sheet.

An accurate description of charge transport in large area graphene and in  GNRs can be obtained by solving in the physical space the semiclassical Boltzmann equations \rev{(BE) \cite{markowich2012semiconductor,jungel2009transport} -- also called quantum Boltzmann equations --} for charge transport by resorting to Direct Simulation Monte Carlo (DSMC) \cite{RoMajCo} or deterministic approach as WENO scheme \cite{LMS} or discontinuous Galerkin (DG) methods \cite{Maj,Majo,CoNa,MajNaRo,ART:NaCaRo_cicp,Battiato1,Battiato2,NaBoRo}.  Other approaches present in the literature include the adoption of drift-diffusion models, hydrodynamical models \cite{Bar,CaRo,LuRo2,MaRo1} (for a comprehensive review see \cite{Bookcamaro}) or the Wigner equation \cite{Mor,Mus,MuWa,CaRoVi}. 

\rev{In order to reduce the computational complexity of the problem, we have adopted an effective model with reduced dimensionality obtained by integrating with respect to the transversal direction ($y$-component in Fig. \ref{fig:GNR}) and, as also done in \cite{ART:Katsnelson}, reformulating the effect of the edge as a further scattering. After the homogenization with respect to the $y$-variable, since the material is homogeneous with respect to the longitudinal direction as well, also the dependence on the $x$-variable is dropped.  We remark that the effect of the finite width of the nanoribbon is still present in the effectively adopted model as a scattering term and as a parameter in the energy dispersion relation.}

Despite the very huge literature on these topics, to quantify the impact of deviations from the classical deterministic modelling setting to include uncertain geometries and molecular mechanics properties of the materials, new approaches based on uncertainty quantification (UQ) methods should be developed. Indeed, to ensure performance reliability it is important to develop a predictive framework that is robust with respect to inevitable fabrication errors and initial conditions parametrizing the molecular dynamics that are based on nanoscale experiments. In the present paper we will tackle issues of  uncertainty quantification related to the electrical properties of graphene but a similar analysis can be performed  for the mechanical features as well, see e.g. \cite{Minh, escalante2022stochastic}.   

The impact of uncertainties in the parameters of a physical model implies additional computational challenges as it increases the dimensionality of the problem in view of the presence of random inputs in the model parameters and initial conditions. This problem is particularly important in kinetic-type models that are already high-dimensional in the physical space, see e.g \cite{jinpareschi,pareschi21} in related modelling settings. In recent years, several UQ methods have been developed to deal with the emerging challenges in the direction of understanding the propagation of missing information and approximating with high accuracy the evolution of quantities of interest (QoI). In this direction,  stochastic Galerkin (sG) methods have been developed to achieve spectral accuracy in the random space for smooth solutions, see e.g.  \cite{HJ,JXZ}. However, in contrast to stochastic collocation methods \cite{DP19}, a direct application of sG methods may lead to the loss of structural properties, like the conservation of large time distributions, non negativity and hyperbolicity in the fluid regime, see e.g.  \cite{DPL,DPZ24,GvdSVK,Poette,Zanella20}. To address these issues, an alternative approach based on the sG reconstruction of the trajectories of colliding particles has been proposed in \cite{CaZa,CaPaZa,ParZan} and further generalized in \cite{BCMZ_Landau,MePaZan,MePaZan_LFP}. We remark that previous results on sG scheme that avoid the loss of hyperbolicity have been based on suitable modifications of  polynomial expansions in a finite volume setting \cite{DEN,GHI}.  

In this article, we focus on the extension of these methods for the semiclassical Boltzmann equation for charged transport in GNRs in the presence of uncertainties. To this end, we develop an efficient particle scheme in the absence of uncertainties and then, after a suitable generalization of the scheme to the case of random inputs, we project the particle dynamics into the space of orthogonal polynomials from which kinetic distributions can be reconstructed. Therefore, the particle-based scheme will preserve the physical properties and the positivity of the distribution function also in the presence of a complex scattering describing the transport of electrons.   

The rest of the manuscript is organized as follows. In Section \ref{SEC:BE} we introduce the semiclassical Boltzmann equation and in Section \ref{sect:num_det} we discuss a particle scheme for transport-collision where the collision step is solved by means of a classical DSMC approach. Then in Section \ref{sect:semiclassical}  a reformulation of the particle scheme by means of sG  methods is given. Finally, in Section \ref{sec:numerics}  several numerical results to show the effectiveness of the approach are presented and discussed. 

\section{Semiclassical BE for charge transport in GNRs} \label{SEC:BE}
\begin{figure}[ht]
\centering
\includegraphics[width = 0.5\linewidth]{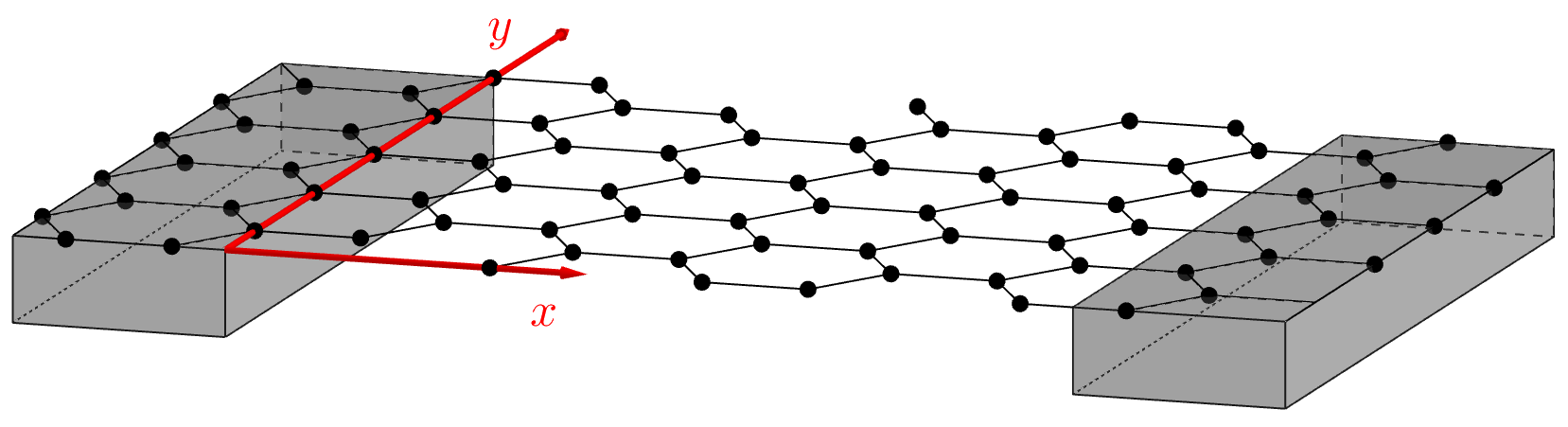}
\caption{\small{Schematic representation of a graphene nanoribbon. Note the irregular edges.}}
\label{fig:GNR}
\end{figure}

An accurate description of charge transport for electrons in the conduction band of GNRs is given by the semiclassical Boltzmann equation in space dimension $2$
\begin{equation}
	\frac{\partial f(t,\bx,\bk)}{\partial t} + \mathbf{v}(\bk)\cdot\nabla_{\bx} f(t,\bx,\bk) - \frac{e}{\hbar}\mathbf{E}(t,\bx)\cdot\nabla_{\bk} f(t,\bx,\bk) = \mathcal{C}[f,f](t,\bx,\bk),
\end{equation}
where $f(t,\bx,\bk)$ is the distribution of electrons at time $t>0$, position $\bx\in\Omega\subseteq\R^2$ and wave-vector $\bk =(k_x,k_y) \in\mathcal{B}\subseteq\R^2$. $\Omega$ is the domain representing the nanoribbons  (see Fig. \ref{fig:GNR}) while  
$\mathcal{B}$ is the first Brillouin zone. The electric field $\mathbf{E}(t,\bx)$ is considered as external and $e$ represents the positive elementary charge.

The group velocity $\mathbf{v}\in\R^2$ is defined as
\begin{equation}
	\mathbf{v}(\bk)=\frac{1}{\hbar}\nabla_{\bk}\varepsilon(\bk).
\end{equation}
where $\varepsilon(\bk)$ is the energy band. In nanoribbons the spacial confinement of the carriers induces an energy gap. We adopt the description proposed in \cite{ART:Bresciani}, according to which the dispersion relation is given by
$$
\varepsilon(\bk) = \hbar v_F \sqrt{k_x^2+k_y^2+\left(\frac{\pi}{W}\right)^2}
$$
with \rev{$\bk$ belonging to $\R^2$.} Here,
$\hbar$ is the reduced Planck's constant, $v_F$ is the Fermi velocity in graphene, and $W$ is the width of the GNR strip, which is a crucial parameter for the quantification of the electric properties. This model is in good agreement with \rev{Density Function Theory (DFT)} up to energies of one eV \cite{ART:Bresciani}. \rev{In principle the wave-vector belongs to the first Brillouin zone $\mathcal{B}$ which is a compact set. However, only a restricted region of $\mathcal{B}$ is indeed populated. They are neighbourhoods (usually called valleys in solid state physics) of the local minima of the energy bands. So, it is a standard approximation, see for example \cite{BOOK:Jacoboni}, to split the electron populations in several subpopulations, one for each valley, and extend each valley to all $\R^d$ (physically $d=2,3$). The rationale is that the occupation number of the charge carriers for all the semiconductor materials is decreasing very fast with energy measured from the minimum. This can be seen from the full band dispersion relation obtained with intensive ab initio calculations \cite{CaNe}. In particular, for graphene the points of minimum of the conduction bands (which are also points of maximum for the valence bands) are just the vertices of the boundary of $\mathcal{B}$, the so-called Dirac points. Therefore, the standard approximation made in the simulation of charge transport in graphene is to consider a population around each Dirac point, virtually extending the neighbourhood to all $\R^2$. This indeed involves parts of three adjacent Brillouin zones}. 

In order to prevent a relevant current due to the electrons in the valence band, a width between 5 nm and 15 nm should be considered \cite{ART:NaCaRo_cicp}. In this case, it is fully justified to consider the semiclassical Boltzmann equation for charge transport in the conduction band only while, in general, the transport equations for both valence and conduction bands must be considered. In the sequel  we will investigate cases where the current due to the electrons in the valence band is negligible.

In the following, we are interested in the construction of a novel Monte Carlo method for the collisional operator $\mathcal{C}[f,f]$; for this reason we first concentrate on the space homogeneous case \cite{MajNaRo,ART:NaCaRo_cicp}. In such a situation the BE reads
\begin{equation}
	\frac{\partial f(t,\bk)}{\partial t} - \frac{e}{\hbar}\mathbf{E}\cdot\nabla_\bk f(t,\bk) = \mathcal{C}[f,f](t,\bk),
	\label{eq:Boltzmann}
\end{equation}
where the electric field $\mathbf{E}=(E_x,E_y)$ is assumed to be  constant in time as well. 
The equation \eqref{eq:Boltzmann} is \rev{an integral-differential} equation in dimension 2 with respect to the $\bk$-space and must be complemented by the  initial conditions which are assumed to be given by a Fermi-Dirac distribution
$$
f(0,\bk) = \frac{1}{1+\exp\left[ \frac{\eps(\bk)-\eps_F}{k_B T} \right]},
$$
where $\varepsilon_F$ is the Fermi level, \rev{and $k_B$ is the Boltzmann constant.}

The collision term $\mathcal{C}[f,f]$ includes both the interaction of electrons with the phonons of graphene and an additional scattering modelling the edge effects. The general form of $\mathcal{C}$ consists of the sum of several contribution
\begin{equation}
	\begin{aligned}
		\mathcal{C}[f,f](t,\bk) & = \int_{\R^2} \sum_{\nu} S^{(\nu)}(\bk',\bk)f(t,\bk')(1-f(t,\bk))\\
		& - \int_{\R^2} \sum_{\nu} S^{(\nu)}(\bk,\bk')f(t,\bk)(1-f(t,\bk')),
	\end{aligned}
\end{equation}
where $\nu$ labels the specific scattering. In graphene there are the acoustic scattering (AC) and three relevant optical phonon scatterings, i.e. the longitudinal optical (LO), the transversal optical (TO) and K phonons (K). In this case, the transition rate $S^{(\nu)}(\bk',\bk)$ is of the form
\begin{equation}
	\begin{aligned}
		S^{(\nu)}(\bk', \bk) = \left| M^{(\nu)}(\bk', \bk) \right|^{2} & \left[ 
		\left( n^{(\nu)}_{\mathbf{q}} + 1 \right) \delta \left( \eps(\bk) - \eps(\bk') + \hbar \, \omega^{(\nu)}_{\mathbf{q}} \right) \right.\\
		& \left. + n^{(\lambda)}_{\mathbf{q}} \, \delta \left( \eps(\bk) - \eps(\bk') - \hbar \, \omega^{(\nu)}_{\mathbf{q}} \right) \right] .
	\end{aligned}
\end{equation}
$\left| M^{(\nu)}(\bk', \bk) \right|$ represents the matrix element of the scattering due to the phonons of type $\nu$ \cite{Borysenko,Barry}. The symbol $\delta$ denotes the Dirac distribution, $\omega^{(\nu)}_{\mathbf{q}}$ is the the $\nu$-th  phonon  frequency, $n^{(\nu)}_{\mathbf{q}}$ is the Bose-Einstein distribution for the phonon of type $\lambda$
$$
n^{(\lambda)}_{\mathbf{q}} = \left[ \exp\left( \frac{\hbar \omega^{(\lambda)}_{\mathbf{q}}}{k_B T_L} \right)-1 \right]^{-1},
$$
where $T_L$ is the lattice temperature, assumed constant in this article, which means that phonons are assumed as a thermal bath. 

The scattering with the acoustic phonons is  considered in the elastic approximation
\begin{equation}
	2 \, n^{(AC)}_{\mathbf{q}}
	\left| M^{(AC)}(\bk', \bk) \right|^{2} = \frac{2}{(2\pi)^2} \dfrac{\pi \, D_{AC}^{2} \, k_{B} \, T_L}{4 \hbar \, \sigma_m \, v_{p}^{2}}
	\left( 1 + \cos \vartheta_{\bk, \bk'} \right) ,
	\label{transport_acoustic}
\end{equation}
where $D_{AC}$ is the acoustic phonon coupling constant, $v_{p}$ is the sound speed in graphene, $\sigma_m$ the graphene areal density, and $\vartheta_{\bk , \bk'}$ is the convex angle between $\bk$ and ${\bk'}$.

For the longitudinal and transversal optical scattering usually is convenient to consider their sum, that is
\begin{equation}\label{transport_optical}
	 \rev{ \left| M^{(O)}(\bk', \bk) \right|^{2} =\left| M^{(LO)}(\bk', \bk) \right|^{2} + \left| M^{(TO)}(\bk', \bk) \right|^{2} =
	\frac{2}{(2\pi^2)} \, \dfrac{\pi \, D_{O}^{2}}{\sigma_m \, \omega_{O}},}
\end{equation}
where $D_{O}$ is the optical phonon coupling constant, $\omega_{O}$ the optical phonon frequency, 

The matrix element of the scattering with K-phonons is
\begin{equation}\label{transport_K}
	\left| M^{(K)}(\bk', \bk) \right|^{2} = 
	\frac{2}{(2\pi)^2}  \dfrac{\pi \, D_{K}^{2}}{\sigma_m \, \omega_{K}}
	\left( 1 - \cos \vartheta_{\bk , \bk'} \right) ,
\end{equation}
where $D_{K}$ is the K-phonon coupling constant and $\omega_{K}$ the K-phonon frequency.

\rev{We remind that a scattering leaves an electron in the same valley (intra-valley scattering) or moves it to an adjacent valley (inter-valley scattering). Moreover, after a scattering an electron in the conduction band can remain in the same band or to the valence one. In this paper only intra-valley and intra-band scattering will be considered. }

If the nanoribbon is on an oxide, e.g. SiO$_2$, there is the additional scattering with the phonons of the substrate whose main effect is a degradation of the mobility \cite{NaRo_CNSNS,CoNa,NaRo_CAIM}. Here only  suspended nanoribbons will be investigated. 

A major difference with respect to the larger area graphene is the edge effect which in principle enters as a boundary condition.  
 In~\cite{ART:Katsnelson} the edges roughness is described by adopting the Berry-Mondragon (or infinite-mass) boundary condition. It corresponds to the single Dirac cone approximation and therefore is applicable for smooth enough disorder near the edges. An alternative (more practical) way to include the edge roughness in the BE is to consider it as an additional scattering  in the collision terms~\cite{ART:Katsnelson}. Within such an approach
the transition rate due to the scattering between the electron and the edge (el-edg) is given by
\begin{equation}\label{transport_el_edg}
	S^{\mathrm{(el-edg)}}(\bk',\bk) = \frac{2}{(2\pi)^2} \frac{\pi N_i V_0^2}{\hbar W} \exp(-2(k_x-k_x')^2 a^2)\delta(\eps(\bk)-\eps(\bk')),
\end{equation}
where $N_i$ is the linear density of defects along the graphene edge and the quantity
\begin{equation}
	V_0\exp(-(k_x-k_x')^2 a^2)
\end{equation}
is the matrix element due to the potential of a single scattering at the edge, being $V_0$ a constant and $a$ a characteristic range, along the edge direction, so that electron scattering with rather strong  longitudinal momentum transfer, this is along the $x$-direction with $\vert k_x-k_x' \vert >1/a$, is effectively suppressed.

\section{Particle-based methods in the deterministic setting}\label{sect:num_det}
In this section we concentrate on the numerical approximation of \eqref{eq:Boltzmann} in the absence of uncertainties. Without pretending to revise the vast literature on numerical methods for the semiclassical Boltzmann equation, we mention discontinuous Galerkin methods \cite{Maj,Majo,MajNaRo} and particle methods belonging to the class of direct simulation Monte Carlo methods (DSMC), see e.g. \cite{RoMajCo,ART:NaCaRo_cicp,BT98} and the references therein. We are interested in the evolution of the density $f(t,\bk)$, $\bk\in \mathbb R^2$, $t\ge0$, solution to \eqref{eq:Boltzmann} complemented with initial condition $f(0,\bk) = f_0(\bk)$.
For the numerical resolution of \eqref{eq:Boltzmann} we adopt the scheme introduced in \cite{RoMajCo}. We consider a time discretization of the interval $[0,T]$ of step $\Delta t>0$, such that $t^n=n\Delta t$. We denote by $f^n(\bk)$ an approximation of $f(t^n,\bk)$ at the $n$-th time step and we perform a splitting method between the electric forcing term and the collisions \cite{MePaZan,dimarco2015numerical}. First we solve the electric forcing term step $f^*=\mathcal{T}_{\Delta t}(f^n)$
\begin{equation}\label{eq:f_star}
\left\lbrace
\begin{aligned}
&\frac{\partial f^*}{\partial t} -\frac{e}{\hbar}\mathbf{E}\cdot\nabla_\bk  f^* = 0\\
&f^*(0,\bk) = f^n(\bk)
\end{aligned}
\right.
\end{equation}
and then we solve the collision step $f^{**}=\tilde{\mathcal{C}}_{\Delta t}(f^*)$
\begin{equation}\label{eq:f_star_star}
\left\lbrace
\begin{aligned}
&\frac{\partial f^{**}}{\partial t}  = \mathcal{C}[f^{**},f^{**}]\\
&f^{**}(0,\bk) = f^*(\Delta t,\bk).
\end{aligned}
\right.
\end{equation}
The first order time splitting finally reads 
\[
f^{n+1}(\bk)=\tilde{\mathcal{C}}_{\Delta t}(\mathcal{T}_{\Delta t}(f^n)(\bk)).
\]
Higher order time splittings are possible, see e.g. \cite{MePaZan,dimarco2015numerical} and the references therein. In the following, we will address the two steps separately.

To numerically solve the electric forcing term and the collision step, we introduce an approximation of the distribution function with a sample of $N$ particles identified by their wave-vectors $\bk_i^n$ at the time $t^n$, for $i=1,2,\ldots,N$, such that
$$
f^n(\bk) \approx f^n_N(\bk) = \sum_{i=1}^N \zeta_i \delta(\bk-\bk^n_i),
$$
where $\zeta_i>0$ are the weights of the particles and $\delta(\cdot)$ the Dirac delta.

We fix an upper and lower bound for the $\bk$-space, to restrict our computational domain $[-L_k,L_k]\times [-L_k,L_k]$ and discretize it in $N_k\times N_k$ uniform cells $C_{rl}$ of size $\vert C_{rl}\vert = \Delta k^2$. 
\rev{We observe that just for moderate values of the wave-vector measured from the Dirac point the distribution is practically zero. This justifies the assumption that the numerical domain is a compact set with zero boundary conditions on the wave-vector. An analytical justification can be also deduced from the appendix to the paper \cite{NaBoRo} where it is proved that the solution fast decay with respect to $\bk$ provided such a condition is met by the initial datum.
To reconstruct the distribution function, several choices are possible. We can apply suitable mollifiers to approximate the Dirac deltas, or the weighted rule, in which each particle belongs to a computational cell and the neighbouring ones with a fraction proportional to the area of overlap. In this work, we reconstruct the distribution $f^n_N$ by histograms counting the number of particles belonging to each cell, which corresponds to spline of order zero}. If we define the occupation number $N_{rl}$ as
$$
N_{rl} = \sum_{i=1}^N \chi(\bk_i^n\in C_{rl}), 	\qquad \forall \, r,l=1,2,\ldots,N_k,
$$
where $\chi(\cdot)$ is the indicator function, \rev{then} the discrete distribution function reads
$$
f^n_{N,rl} = \frac{N_{rl}}{\max_{rl} N_{rl}}, \qquad \forall \, r,l=1,2,\ldots,N_k,
$$
where we have normalized the distribution with its maximum.

Now we introduce the relevant observable macroscopic quantities $\mathcal{O}[f](t)$, which are typical moments or functions of the distribution $f(t,\bk)$ of the type
\[
\mathcal{O}[f](t) = \int_{\R^2} \phi(\bk) f(t,\bk) d\bk,
\]
for some $\phi(\bk)$. For the proposed model, we will consider the electron density $n$, the current density $\mathbf{J}$, the mean velocity $\mathbf{V}$ and the mean energy $\mathcal{E}$. These observables are defined as
\begin{align}
n & = \frac{g_s g_v}{(2\pi)^2}\int_{\R^2} f(0,\bk) \, d\bk,\label{eq:den} \\ 
\mathbf{V}(t) & = \frac{g_s g_v}{(2\pi)^2} \frac{1}{n} \int_{\R^2} \mathbf{v}(\bk) f(t,\bk) \, d\bk \approx \mathbf{V}^n = \frac{1}{N} \sum_{i=1}^N \mathbf{v}(\bk_i^n), \label{eq:Vx}\\
\mathbf{J}(t) & = -en\mathbf{V}(t) \approx \mathbf{J}^n = -en\mathbf{V}^n,\\
\mathcal{E}(t) & = \frac{g_s g_v}{(2\pi)^2} \frac{1}{n} \int_{\R^2} \varepsilon(\bk) f(t,\bk) \, d\bk \approx \mathcal{E}^n = \frac{1}{N}\sum_{i=1}^N \varepsilon(\bk_i^n). \label{eq:en}
\end{align}
In \eqref{eq:den} the factors $g_s$ and $g_v$ represent the spin and the valley degeneracies respectively. We consider $g_s=g_v=2$. The factor $1/(2\pi)^2$ is due to the homogenization \cite{BOOK:Jacoboni}. We remark that the electron density is constant in time because we consider the unipolar case only and it depends on the parameters $\varepsilon_F$ and $W$.

\paragraph{Electric forcing term}
The electric forcing term \eqref{eq:f_star} is solved analytically performing a free-flight step to all the particles according to the semiclassical equation of motion. Consequently, for any wave-vector $i$ we have
$$
\frac{d\bk_i(t)}{dt} = -\frac{e}{\hbar}\mathbf{E}.
$$
At the time discrete level, introducing a first order Euler scheme for the time derivative, we obtain
\be \label{eq:transport}
\bk_i^{n+1} = \bk_i^n - \frac{e}{\hbar}\mathbf{E}\Delta t.
\ee

In this step, in order to preserve the occupation number of the cells, we adopt a Lagrangian approach and move the grid of the same quantity of the variation of the particle wave-vectors \cite{RoMajCo}.

\paragraph{Collision step}
The collision step \eqref{eq:f_star_star} is solved by a direct simulation Monte Carlo method \cite{BOOK:Jacoboni}. For each particle $i$, with $i=1,2,\ldots,N$, the time elapsed between two collisions is determined in a random way according to an exponential distribution, that is
\begin{equation} \label{eq:dt}
	\delta t_i = -\frac{\log \xi_i}{\Gamma^{tot}_i},
\end{equation}
where $\xi_i \sim \mathcal{U}([0,1])$ and $\Gamma^{tot}_i$ is the total scattering rate defined as
\begin{equation}\label{eq:gamma_tot}
\Gamma^{tot}_i\coloneqq \Gamma^{tot}(\bk_i) = (1+\alpha) \sum_{\nu} \Gamma^{\nu}(\bk_i),
\end{equation}
where $\nu$ labels the $\nu$-th phonon mode and $0 < \alpha < 1$ is a tuning parameter. The general expression of $\Gamma^{\nu}(\bk_i)$ is
$$
\Gamma^{\nu}_i\coloneqq\Gamma^{\nu}(\bk_i) = \int_{\R^2} S^{\nu}(\bk_i,\bk') \, d\bk',
$$
see Appendix \ref{App_sc_rate} for further details. Moreover, we introduce the self-scattering rate
$$
\Gamma^{ss}_i\coloneqq\Gamma^{ss}(\bk_i) = \Gamma^{tot}(\bk_i) - \sum_{\nu} \Gamma^{\nu}(\bk_i) = \alpha \sum_{\nu} \Gamma^{\nu}(\bk_i),
$$
which is the scattering rate associated to a fictitious scattering that does not change the state of the electron.

Once $\delta t_i$ is determined for the particle $i$, one of the previously introduced scattering must be selected according to the scattering rates. We distinguish seven possibilities for this choice due to the fact that for optical and K phonons the final state is different if an emission ($-$) or an absorption ($+$) happens. Therefore, we indicate by $\Lambda_i$ the vector which components are all the scattering rates associated with the particle $i$ in an arbitrary order,
\begin{equation} \label{eq:lambda}
\Lambda_i = (\Lambda^m_i)_{m=1}^7 = (\Gamma^{AC}_i,\Gamma^{OP^-}_i,\Gamma^{OP^+}_i,\Gamma^{K^-}_i,\Gamma^{K^+}_i,\Gamma^{el-edg}_i,\Gamma^{ss}_i).
\end{equation}

The final state at a certain time is determined in a random way according to the distribution of the specific selected scattering. Since the Pauli exclusion principle is present, the final state can be accepted only after a rejection on the current particles distribution. This should be done for each transition making the procedure not feasible to be computed in parallel. In this paper, we make a slight approximation in order to overcome the problem considering an acceptance-rejection procedure corresponding to the Pauli exclusion principle on the particles distribution at the previous time step. In Section \ref{sec:numerics} we investigate this approximation with several numerical tests.

In this way, it is possible to reformulate the DSMC method introduced in the literature in a way that all the selection procedures are embedded into the elementary variation of the wave-vector, following the ideas introduced in \cite{ParZan, MeToZan} and subsequent works. To this aim, we determine the candidate update of any wave-vector
\begin{equation}\label{eq:update}
\bk_i' = \sum_{\ell=1}^{7} \bigg[ \chi(s_i\in[G^{\ell-1}_i, G^\ell_i]) \cdot \mathbf{h}^{\ell}_i \bigg]
\end{equation}
and we impose the Pauli principle
\begin{equation} \label{eq:Pauli}
\mathbf{k}^{n+t_c}_i = \chi\bigg(f_{N,rl}^n \geq \eta_i\bigg) \cdot \mathbf{k}^{n}_i + \chi\bigg(f_{N,rl}^n < \eta_i\bigg)\cdot\bk_i'
\end{equation}
where $r$ and $l$ are determined such that $\bk_i'\in C_{rl}$. In the previous expressions, $t_c>0$ is a time counter, $\eta_i$ and $s_i$ are uniform random numbers in the interval $(0,1)$ associated with the particle $i$, and $f^n_{N,rl}$ is the value of the discrete distribution at the previous time in the cell $C_{rl}$. The vector $G_i$ is such that its components are the normalized cumulative sum of the components of $\Lambda_i$, namely
\[
G^0_i = 0, \qquad G^\ell_i = \sum_{m=1}^{\ell} \frac{\Lambda^{m}_i}{\Gamma^{tot}_i}, \qquad \text{with} \qquad \ell=1,\dots,7.
\]
Finally, the $\mathbf{h}^{\ell}_i = \mathbf{h}^{\ell} (\mathbf{k}^{n}_i)$, with $\ell=1,\dots,7$, perform the update according to the selected scattering. In the following, we illustrate in details all the mechanisms to determine the final states. Defining $\theta^{in}$ the angle that the incoming particle of wave-vector $\bk_i^n$ forms with the $k_x$ axis, we have the following possibilities.
\begin{itemize}
\item The $AC$ scattering, corresponding to the index $\ell=1$,
\rev{\[
\mathbf{h}_1(\mathbf{k}_i^{n})=\mathbf{h}^{ac}(\mathbf{k}_i^{n})=|\mathbf{k}_i^n| 
\begin{bmatrix}
	\cos \theta^{ac} \cos \theta^{in} - \sin \theta^{ac} \sin \theta^{in} \\ \sin \theta^{ac} \cos \theta^{in} + \cos \theta^{ac} \sin \theta^{in}
\end{bmatrix},
\]}
where $\theta^{ac}$ is sampled according to the distribution 
\[
\theta^{ac} \sim 1+\cos\theta
\]
which follows from \eqref{transport_acoustic}.
\item The $OP$ scattering in emission ($-$) and absorption ($+$), corresponding to the indices $\ell=2,3$,
\rev{\[
\mathbf{h}^{OP^\pm}(\mathbf{k}_i^{n})=\left(|\mathbf{k}_i^n|\pm \frac{\omega_{OP}}{v_F}\right) 
\begin{bmatrix}
	\cos \theta^{OP} \cos \theta^{in} - \sin \theta^{OP} \sin \theta^{in} \\ \sin \theta^{OP} \cos \theta^{in} + \cos \theta^{OP} \sin \theta^{in}
\end{bmatrix},
\]}
in a way that $\mathbf{h}_{2}=\mathbf{h}^{OP^-}$, $\mathbf{h}_{3}=\mathbf{h}^{OP^+}$. In the above expression, $\theta^{OP}$ is uniformly distributed in $(0,2\pi)$ 
\[
\theta^{OP} \sim \mathcal{U}([0,2\pi])
\]
according to (\ref{transport_optical}).
\item The $K$ scattering in emission ($-$) and absorption ($+$), corresponding to the indices $\ell=4,5$,
\rev{\[
\mathbf{h}^{K^\pm}(\mathbf{k}_i^{n})=\left(|\mathbf{k}_i^n|\pm \frac{\omega_{K}}{v_F}\right) 
\begin{bmatrix}
	\cos \theta^{K} \cos \theta^{in} - \sin \theta^{K} \sin \theta^{in} \\ \sin \theta^{K} \cos \theta^{in} + \cos \theta^{K} \sin \theta^{in}
\end{bmatrix},
\]}
with $\mathbf{h}_{4}=\mathbf{h}^{K^-}$, $\mathbf{h}_{5}=\mathbf{h}^{K^+}$. Here, $\theta^{K}$ is determined in a random way according to the distribution
\[
\theta^{K} \sim  1-\cos\theta
\]
which follows from \eqref{transport_acoustic}.
\item The  $el-edg$ scattering, corresponding to the index $\ell=6$,
\rev{\[
\mathbf{h}_6(\mathbf{k}_i^{n})=\mathbf{h}^{el-edg}(\mathbf{k}_i^{n})=|\mathbf{k}_i^n| 
\begin{bmatrix}
	\cos \theta^{el-edg} \cos \theta^{in} - \sin \theta^{el-edg} \sin \theta^{in} \\ \sin \theta^{el-edg} \cos \theta^{in} + \cos \theta^{el-edg} \sin \theta^{in}
\end{bmatrix}.
\]}
In this case, the quantity $\theta^{el-edg}$ is determined in a random way according to the distribution 
$$
\theta^{el-edg} \sim \exp(-2\vert\bk_i^n\vert^2(\cos\theta^{in}-\cos\theta^{el-edg})^2 a^2),
$$
which follows from \eqref{transport_el_edg}.
\item The self scattering, corresponding to the index $\ell=7$, which reads 
$$
\mathbf{h}_7(\mathbf{k}^n_i)=\mathbf{k}^n_i.
$$
\end{itemize}
To summarize, the algorithm reads:

\begin{minipage}{.9\linewidth}
\begin{algorithm}[H]  
\footnotesize
\caption{\small{Collision step} }
\label{alg:collision:det}
\begin{itemize}
	\item for $n=0,\dots,T_f/\Delta t-1$, given $f^{n}_N$:
	\begin{itemize}
		\item for $i=1,2,\dots,N$, given $\mathbf{k}^{n}_{i}$:
		\begin{itemize}
			\item set the time counter $t_c=0$; 
			\item compute the time step $\delta t_i$ according to \eqref{eq:dt};
			\item while $t_c+\delta t_i<\Delta t$:
			\begin{itemize}
				\item compute $\mathbf{k}'_{i}$ according to \eqref{eq:update};
				\item compute $\mathbf{k}^{n+t_c}_{i}$ according to \eqref{eq:Pauli};
				\item $t_c = t_c + \delta t_i$;
			\end{itemize}
			\item end while;
			\item set $\mathbf{k}^{n+1}_{i}=\mathbf{k}^{n+t_c}_{i}$;
		\end{itemize}
		\item end for;
	\end{itemize}
	\item end for.
\end{itemize}
\end{algorithm}
\end{minipage}

\begin{remark}
In the Algorithm \ref{alg:collision:det}, the for loop on the particle index $i$ can be parallelized. This is possible because the rejection due to the Pauli principle is performed with the distribution function of the previous time step. Consequently, \rev{it is not necessary} to compute the distribution for each scattering of each particle. In this way, there is a significant reduction of the computational cost. Another motivation is due to the possibility to adopt the particle sG method. It will be clarified in the next section. 
\end{remark}
\begin{remark}
The consistency with the Pauli principle is not guaranteed with the above introduced approximation on the distribution function, differently from \cite{RoMajCo}. 
The error stemming from the proposed scheme will be numerically evaluated in the next section.
\rev{Discussions on the violation of the Pauli exclusion principle for different Monte Carlo algorithms have been made in \cite{coco2021pauli} and the references therein}
\end{remark}

\section{Semiclassical BE in the presence of uncertainties}\
\label{sect:semiclassical}
In this section we focus on a stochastic Galerkin particle formulation of the proposed scheme. We highlight that these methods have been recently introduced for kinetic equations in classical rarefied gas dynamics, plasma physics and collective phenomena, see e.g.\cite{CaZa,CaPaZa,ParZan,MePaZan,MePaZan_LFP,Poette}.
Now we include in  the space homogeneous semiclassical Boltzmann equation the presence of uncertainties
\[
\frac{\partial }{\partial t} f(t,\bk,\z) - \frac{e}{\hbar}\mathbf{E}(\z)\cdot\nabla_{\bk} f(t,\bk,\z) = \mathcal{C}[f,f](t,\bk,\z),
\]
with uncertain initial condition
$$
f(0,\bk,\z) = \frac{1}{1+\exp\left[ \frac{\eps(\bk,\z)-\eps_F}{k_B T} \right]},
$$
where $\z=(z_1, \dots, z_{d_\z} )\in I_\z \subseteq \R^{d_\z}$ is a vector of random variables of dimension $d_\z\in\mathbb{N}$ collecting all the uncertainties of the system, distributed as $p(\z)$ in a way that
\[
\textrm{Prob}(\z\in I_\z) = \int_{I_\z} p(\z)d\z.
\]
In the following, we reformulate the particle scheme introduced in the previous section to the uncertain scenario. In particular, we consider a generalized Polynomial Chaos (gPC) expansion of the wave-vectors in the space of the random parameters and a subsequent projection of the particle scheme in the linear space of the polynomials. This approach has been introduced in \cite{ParZan} for the Boltzmann equation and in \cite{MePaZan_LFP,MePaZan,BCMZ_Landau} for plasma equations, while in \cite{MeToZan,CaZa,CaPaZa} the reader can find applications to systems with collective behaviour.

\subsection{Stochastic Galerkin projection of the particle scheme}
We consider a sample of $N$ particles identified by their wave-vectors $\bk_i^n(\z)$ at the time step $n$, for $i=1,2,\ldots,N$, depending on the random parameter $\z$. Their gPC expansion truncated at the order $M\in\mathbb{N}$ reads
\be \label{eq_gPC}
\bk_i^n(\z) \approx \bk_i^{n,M}(\z) = \sum_{h=0}^{M} \hat{\bk}^n_{i,h} \Psi_h(\z),
\ee
where $\{\Psi_h(\cdot)\}$, for $h=0,1,\ldots,M$, are polynomials of order $h$ orthonormal with respect to the distribution of the random parameter
\[
\int_{I_{\z}} \Psi_h(\z) \Psi_\ell(\z) p(\z) d\z = \mathbb{E}_{\z}[\Psi_h(\cdot)\Psi_\ell(\cdot)] = \delta_{h\ell},
\]
where $\delta_{h\ell}$ is the Kronecker delta. The coefficients of the expansion are given by the projection of the wave-vectors onto the linear space of the polynomials, namely
\[
\hat{\bk}^n_{i,h} \coloneqq \int_{I_{\z}} \bk^n_{i}(\z) \Psi_h(\z) p(\z) d\z = \mathbb{E}_{\z}[\bk^n_{i}(\cdot)\Psi_h(\cdot)].
\]
The choice of the polynomials depends on the distribution $p(\z)$ and it is done by following the so-called Wiener-Askey scheme \cite{xiu02,xiu10}.

With the approximation of the wave-vectors given by \eqref{eq_gPC}, the discrete distribution function at the time step $n$ reads
\be \label{eq:f_gPC}
\rev{f^{n,M}_{N,rl}(\z) = \frac{N^M_{rl}(\z)}{\max_{rl}N^M_{rl}(\z)},}
\ee
where the uncertain occupation numbers are
\[
N^M_{rl}(\z) = \sum_{i=1}^N \chi(\bk^{n,M}_i \in C_{rl}), \qquad \forall\, r,l=1,2,\ldots,N_k.
\]

\paragraph{Stochastic Galerkin electric forcing term}
The stochastic Galerkin electric forcing term is found first by substituting the gPC expansion of any particle $i$ \eqref{eq_gPC} into \eqref{eq:transport}, which is the discrete solution to \eqref{eq:f_star} 
\[
\bk_i^{n+1,M}(\z) = \bk_i^{n,M}(\z) - \frac{e}{\hbar}  \mathbf{E}(\z) \Delta t. 
\]

Then, we integrate in $I_\z$ against $\Psi_h(\z)p(\z)d\z$ for every $h=0,1,\ldots,M$. Exploiting the orthonormality of the polynomials, we obtain the time evolution of the coefficients of the expansion
\[
\hat{\bk}_{i,h}^{n+1} = \hat{\bk}_{i,h}^n - \frac{e}{\hbar}\Delta t \int_{I_\z} \mathbf{E}(\z) \Psi_h(\z)p(\z)d\z.
\]
We observe that, since the electric field is assumed constant in time, the term inside the integral can be calculated offline, thus increasing the computational performance of the scheme.  

\paragraph{Stochastic Galerkin collision step}
The collision step in the presence of uncertainty is obtained similarly. We first observe that, since $\bk^n_i=\bk^n_i(\z)$, for every $i=1,\ldots,N$, all the scattering rates defining the vector \eqref{eq:lambda} depend on the random parameter, and thus $\Lambda_i=\Lambda_i(\z)$, $G_i=G_i(\z)$. Obviously, we also have $\mathbf{h}^{\ell}_i= \mathbf{h}^{\ell}(\mathbf{k}^{n}_i,\z)$, since $\theta^{in}=\theta^{in}(\z)$ and $\theta^{el-edg}=\theta^{el-edg}(\z)$. We observe that $\theta^{ac}$, $\theta^{OP^{\pm}}$ and $\theta^{K^{\pm}}$ are instead sampled from deterministic distributions.
Moreover, we consider a deterministic time between two collisions by taking the minimum in $\z$ of all the $\delta t_i(\z)$
\be \label{eq:deltat}
\delta \tilde{t}_i = \min_{\z\in I_\z} \delta t_i(\z) = - \frac{\log \xi_i}{\max_{\z\in I_\z} \Gamma^{tot}_i(\z)}.
\ee
\rev{The reason why we need a deterministic $\delta\tilde{t}_i$ is related to the projection of the collision algorithm, which will be presented below. By taking the minimum in $\textbf{z}$ over the $\delta t_i(\textbf{z})$, we select the worst case scenario from a computational efficiency perspective, but the one that is consistent with the physics of the problem. Indeed, the collision process includes the possibility of the self scattering, which does not alter the state of the particles.}

We consider now the gPC expansion of the wave-vectors into the microscopic rule \eqref{eq:update}-\eqref{eq:Pauli} 
\be \label{eq:update_gPC}
\bk_i'^M(\z) = \sum_{\ell=1}^{7} \bigg[ \chi(s_i\in[G^{\ell-1}_i(\z), G^\ell_i(\z)]) \cdot \mathbf{h}^{\ell,M}_i(\z) \bigg]
\ee
\be \label{eq:Pauli_gPC}
\mathbf{k}^{n+t_c,M}_i(\z) = \chi\bigg(f^{n,M}_{N,rl}(\z) \geq \eta_i \bigg) \cdot \mathbf{k}^{n,M}_i(\z) + \chi\bigg(f^{n,M}_{N,rl}(\z) < \eta_i \bigg) \cdot \bk_i'^M(\z),
\ee
where $\mathbf{h}^{\ell,M}_i(\z)\coloneqq \mathbf{h}^{\ell}(\mathbf{k}^{n,M}_i(\z))$ and $f^{n,M}_{N,rl}(\z)$ is the distribution function at the time step $n$ evaluated in the cell $C_{rl}$.

Then we project \eqref{eq:Pauli_gPC} into the linear space of the polynomials by integrating in $I_\z$ against $\Psi_h(\z)p(\z)d\z$ to obtain
\be \label{eq:update_sG}
\begin{split}
\hat{\mathbf{k}}^{n+t_c}_{i,h} = \mathcal{K}^{(1)}_{i,h} +  \mathcal{K}^{(2)}_{i,h}
\end{split}
\ee
with the collisional matrices 
\be \label{eq:M1}
\mathcal{K}^{(1)}_{i,h} = \int_{I_\z} \chi\bigg(f^{n,M}_{N,rl}(\z) \geq \eta_i \bigg) \cdot \mathbf{k}^{n,M}_i(\z) \Psi_h(\z)p(\z)d\z,
\ee
\be \label{eq:M2}
\mathcal{K}^{(2)}_{i,h} =  \int_{I_\z}\chi\bigg(f^{n,M}_{N,rl}(\z) < \eta_i \bigg) \cdot \bk_i'^M(\z) \Psi_h(\z)p(\z)d\z.
\ee
The sG particle algorithm reads:

\begin{minipage}{.9\linewidth}
\begin{algorithm}[H]  
	\footnotesize
	\caption{\small{Stochastic collision step} }
	\label{alg:collision:unc}
	\begin{itemize}
		\item for $n=0,\dots,T_f/\Delta t-1$, given $f^{n,M}_{N,rl}(\z)$:
		\begin{itemize}
			\item for $i=1,2,\dots,N$, given $\hat{\mathbf{k}}^{n}_{i,h}$:
			\begin{itemize}
				\item set the time counter $t_c=0$; 
				\item compute the time step $\delta \tilde{t}_i$ according to \eqref{eq:deltat};
				\item while $t_c+\delta \tilde{t}_i < \Delta t$:
				\begin{itemize}
					\item compute the collisional matrices \eqref{eq:M1}-\eqref{eq:M2};
					\item set $\hat{\mathbf{k}}^{n+t_c}_{i,h}$ according to \eqref{eq:update_sG};
					\item $t_c = t_c + \delta \tilde{t}_i $;
				\end{itemize}
				\item end while;
				\item set $\hat{\mathbf{k}}^{n+1}_{i,h}=\hat{\mathbf{k}}^{n+t_c}_{i,h}$;
			\end{itemize}
			\item end for;
		\end{itemize}
		\item end for.
	\end{itemize}
\end{algorithm}
\end{minipage}

\subsection{Approximation of QoI}
Algorithm \ref{alg:collision:unc} gives the numerical time evolution of the coefficients of the gPC expansion. However, we are interested in understanding the effect of the uncertainty $\z$ on the macroscopic quantities. The observables introduced in the previous section depend now on the random parameter $\mathcal{O}[f](t,\z)$, and thus are uncertain quantities
\begin{align}
	n(\z) & = \frac{g_s g_v}{(2\pi)^2}\int_{\R^2} f(0,\bk,\z) \, d\bk, \\ 
	\mathbf{V}(t,\z) & = \frac{g_s g_v}{(2\pi)^2} \frac{1}{n(\z)} \int_{\R^2} \mathbf{v}(\bk) f(t,\bk,\z) \, d\bk  \approx \mathbf{V}^n(\z) = \frac{1}{N} \sum_{i=1}^N \mathbf{v}(\bk_i^n(\z)), \\
	\mathbf{J}(t,\z) & = -en(\z)\mathbf{V}(t,\z) \approx \mathbf{J}^n(\z) = -e n(\z) \mathbf{V}^n(\z) ,\\
	\mathcal{E}(t,\z) & = \frac{g_s g_v}{(2\pi)^2} \frac{1}{n(\z)} \int_{\R^2} \varepsilon(\bk,\z) f(t,\bk,\z) \, d\bk \approx \mathcal{E}^n(\z) = \frac{1}{N}\sum_{i=1}^N \varepsilon(\bk_i^n(\z)). 
\end{align}	
To obtain their numerical time evolution, once we have every $\hat{\bk}^n_{i,h}$, we first need to reconstruct the particles and the distribution in the physical space according to \eqref{eq_gPC} and \eqref{eq:f_gPC}, and then to compute  
\begin{align} 
	\mathbf{V}^{n,M}(\z)  & = \frac{1}{N} \sum_{i=1}^N \mathbf{v}(\bk_i^{n,M}(\z)), \\
	\mathbf{J}^{n,M}(\z)  & = -e \,n(\z)\,\mathbf{V}^{n,M}(\z),\\
	\mathcal{E}^{n,M}(\z) & = \frac{1}{N}\sum_{i=1}^N \varepsilon(\bk_i^{n,M}(\z)). 
\end{align}
Finally, the so-called Quantities of Interest (QoI) are obtained as expectations with respect to the distribution of the random parameter $\z$ of the macroscopic quantities computed from the reconstructed particles $\bk_i^{n,M}(\z)$ and distribution $f^{n,M}_{N,rl}(\z)$, namely
\[
\mathrm{QoI} = \mathbb{E}_{\z}\left[ \mathcal{O}[f^{n,M}_{N,rl}](\z) \right].
\]

\section{Numerical results} \label{sec:numerics}
In this section, we test the proposed scheme by considering  numerical simulations in GNRs. First, we consider the case without uncertainties in order to investigate the agreement between the standard algorithm and the parallelizable version described in the previous section. Next, we include also the presence of uncertainties. In particular, we will consider two different scenarios corresponding to an uncertain GNR width $W=W(\z)$ and uncertain external electric field $\mathbf{E}=\mathbf{E}(\z)$.
The motivation of these choices is based, on one hand, on an unavoidable inaccuracy in the device fabrication which produces, when large area graphene is cut in ribbons, irregular boundaries \cite{ART:Bresciani} with no negligible effects on the edge scattering. On the other hand, the metallic contacts induce a resistance effect not determinable clearly \cite{xia2011} both for theoretical reasons and for fluctuations in the realization of the metallic-graphene interface.

In all the simulations we adopt for the electron-phonon scatterings and for the electron-edge interactions the physical parameters of Table \ref{TAB:el_ph_edg} \rev{\cite{Borysenko,ART:Katsnelson,Barry}}. We set the  simulation parameters as $\Delta t = 0.0025$ ps, $N_k=240$, $L_k\cdot\hbar v_F=5$ eV, \rev{in a way that $\Delta k \cdot \hbar v_F=2L_k\cdot\hbar v_F/N_k=10/240$ eV}. The tuning parameter for the self scattering in eq. \eqref{eq:gamma_tot} is taken as $\alpha=0.1$. All the other parameters will be specified in the description of the tests.

\begin{table}[!ht]
\centering
\begin{minipage}{.4\linewidth}
\centering
\begin{tabular}{ll}
\hline
Parameter & Value \\
\hline
& \\[-10pt]
$v_{F}$ & $10^{8}$ cm/s \\[2pt]
%
$\sigma_{m}$ & $7.6 \times 10^{-8}$ g/cm$^{2}$ \\[2pt]
%
$\hbar \, \omega_{O}$ & $164.6$ meV \\[2pt]
%
$\hbar \, \omega_{K}$ & $124$ meV \\[2pt]
\hline
\end{tabular}
\end{minipage}
\begin{minipage}{.4\linewidth}
\centering
\begin{tabular}{ll}
\hline
Parameter & Value \\
\hline
& \\[-10pt]
$v_{p}$ & $2 \times 10^{6}$ cm/s \\[2pt]
%
$D_{ac}$ & $6.8$ eV \\[2pt]
%
$D_{O}$ & $10^{9}$ eV/cm \\[2pt]
%
$D_{K}$ & $3.5 \times 10^{8}$ eV/cm \\[2pt]
\hline
\end{tabular}
\end{minipage}

\vspace{0.5cm}


\centering
\begin{tabular}{ll}
\hline
Parameter & Value \\
\hline
& \\[-10pt]
$N_i$ & $10^{4}$ cm$^{-1}$ \\[2pt]
$V_0$ & $4.56 \times 10^{-14}$ eV cm$^{2}$ \\[2pt]
$a$ & $10^{-8}$ cm \\[2pt]
\hline
\end{tabular}
\caption{\small{Physical parameters for the electron-phonon and the electron-edge collision term.}}
\label{TAB:el_ph_edg}
\end{table}

\subsection{Test 1: Semiclassical BE without uncertainties}
In this subsection, we investigate the validity of Algorithm \ref{alg:collision:det} for the semiclassical Boltzmann equation in the absence of uncertainties. We remark that, in this case, the Pauli exclusion principle is checked by performing  a rejection with the distribution at the previous time step. This does not guarantee that the distribution does not become greater than its maximum at the initial time. As a consequence, there is no guarantee that the empirical distribution function could not exceed the maximum occupation number. The latter is fixed for each cell according to the chosen Fermi energy and the number of particles used in the Monte Carlo simulations. We are therefore interested in understanding whether and how often the Pauli exclusion principle is violated. 

To this end, we perform simulations with Algorithm \ref{alg:collision:det} with number of particles $N=10^5,\,10^6$, Fermi energy $\epsilon_F=0.4,\,0.5,\,0.6$ eV, electric field along the $x$-axis $E_x=0.1,\,0.5,\,1$ V/$\mu$m, with $E_y=0$, and GNR width $W=4,\,7,\,10$ nm. We compute the percentage $\bar{N}$ of the highest number of particles for which  the maximum occupation number is violated in the single time step over the time span from 0 to 3 ps. Moreover, we evaluate the percentage relative error of the mean energy $\mathcal{E}(t)$ \eqref{eq:en} and mean velocity $V_x(t)$ \eqref{eq:Vx} with respect to the solution computed with a DSMC scheme consistent with the exclusion Pauli principle \cite{RoMajCo}. We indicate by “Parallel" (P) the numerical solution obtained with Algorithm \ref{alg:collision:det} and by “Non Parallel" (NP) the numerical solution computed with the DSMC introduced in \cite{RoMajCo}. The percentage error of energy and longitudinal component of the electron velocity reads
\[
\textrm{E}_\mathcal{E} = \dfrac{\| \mathcal{E}^{(NP)}(t) - \mathcal{E}^{(P)}(t) \|}{\| \mathcal{E}^{(NP)}(t) \|} \cdot 100,\qquad \textrm{E}_{V_x} = \dfrac{\| V_x^{(NP)}(t) - V_x^{(P)}(t) \|}{\| V_x^{(NP)}(t) \|} \cdot 100.
\]
In Table \ref{tab:confronto5}-\ref{tab:confronto6} we display the results for $N=10^5$ and $N=10^6$, respectively. 
In all the cases the percentage of particles violating Pauli's principle is practically negligible. Therefore, there is a strong numerical evidence that the parallel algorithm performs very well with a considerable reduction of the computational burden. As general considerations the error in the energy increases  by increasing the GNR width while the error in the velocity has the opposite behaviour. The error percentage is always small for the energy but can be about 3\% in the velocity. This could be ascribed to the edge scattering which is elastic and therefore most influential  for the velocity than the energy.  By increasing the number of particles, as expected, the error reduces while the situation becomes a bit worse by raising the Fermi level. 

\begin{figure}
\centering
\includegraphics[width = 0.3\linewidth]{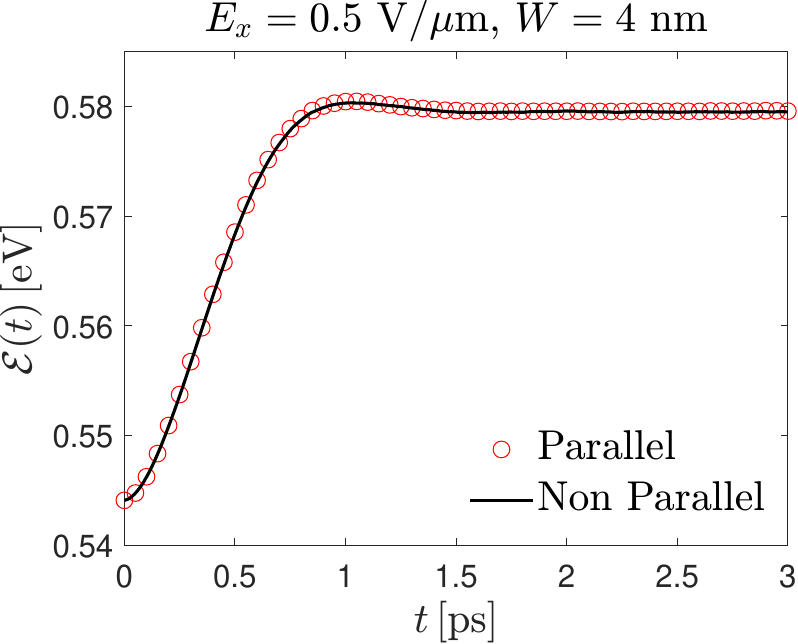}\hspace{2ex}
\includegraphics[width = 0.3\linewidth]{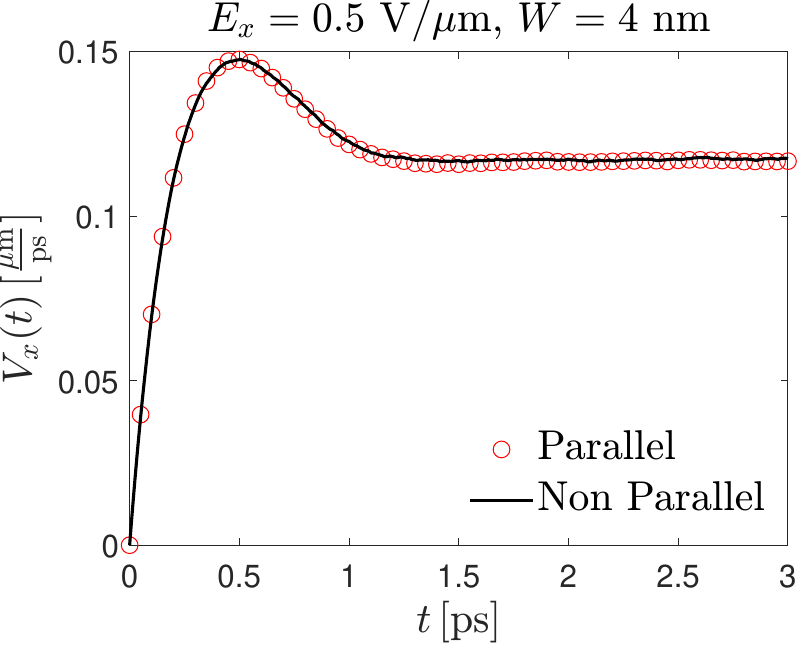} 
\vspace{2ex}

\includegraphics[width = 0.3\linewidth]{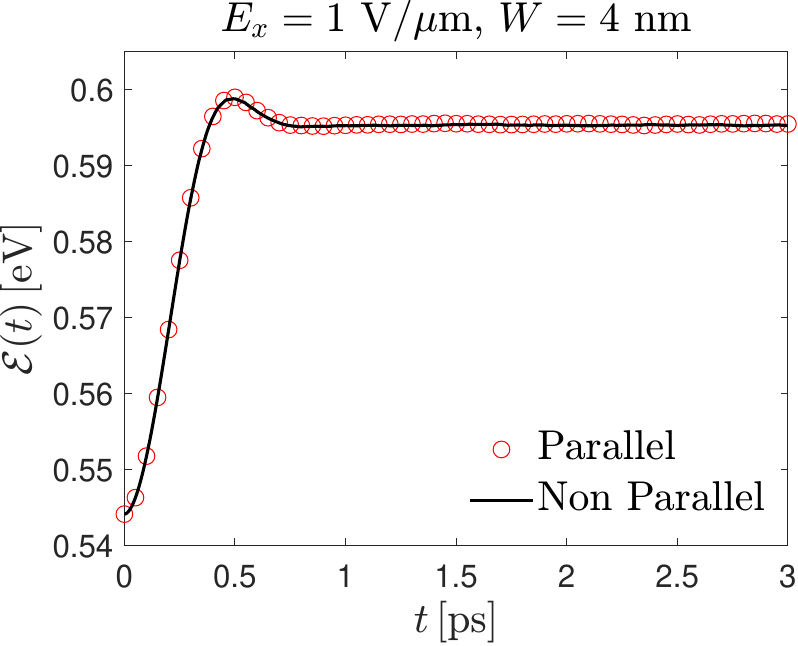}\hspace{2ex}
\includegraphics[width = 0.3\linewidth]{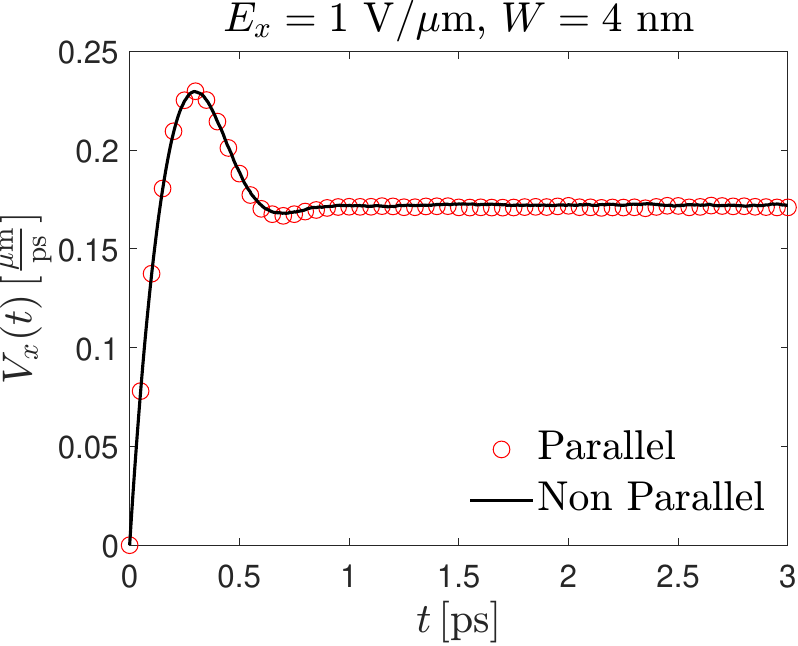}
\vspace{2ex}

\includegraphics[width = 0.3\linewidth]{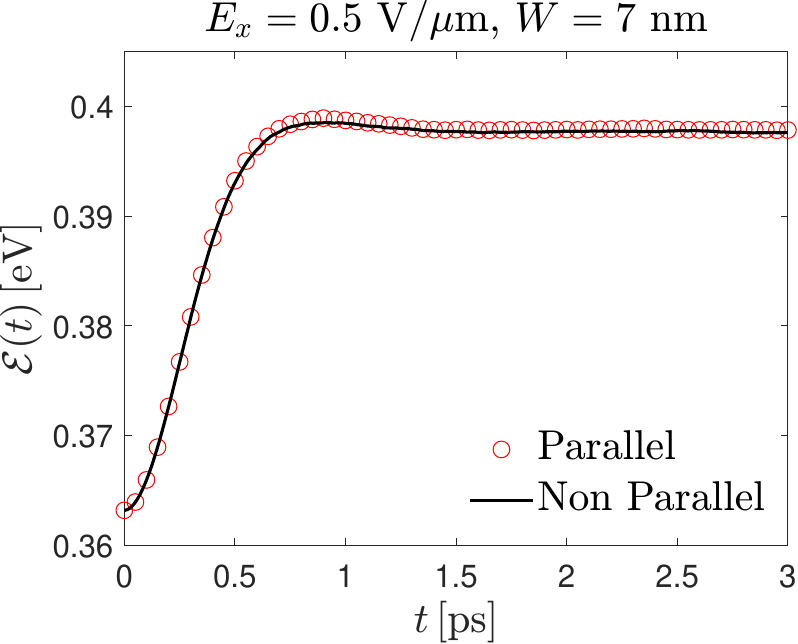}\hspace{2ex}
\includegraphics[width = 0.3\linewidth]{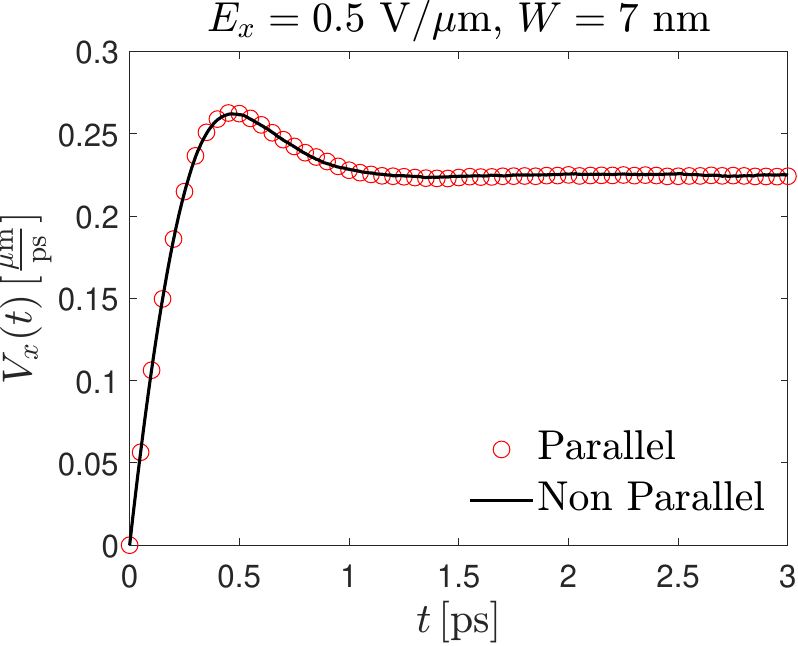} 
\vspace{2ex}

\includegraphics[width = 0.3\linewidth]{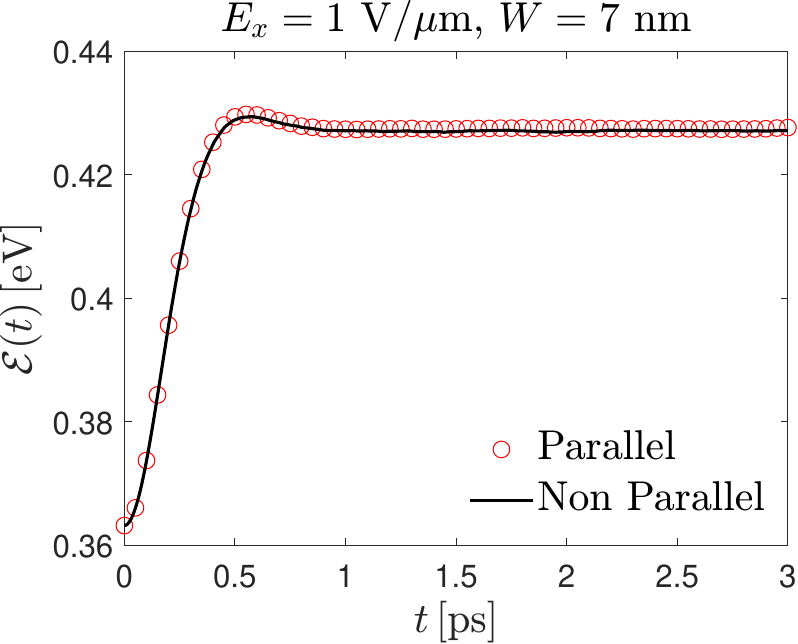}\hspace{2ex}
\includegraphics[width = 0.3\linewidth]{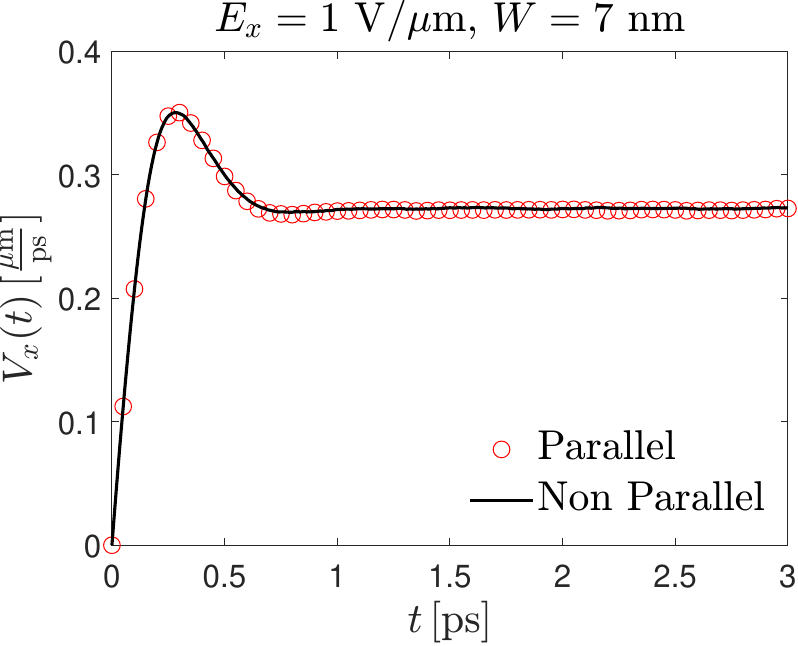}
\caption{\small{\textbf{Test 1:} Comparison between Parallel (red circles) and Non Parallel (solid black line) algorithms of the time evolution of energy $\mathcal{E}(t)$ (left column) and the longitudinal component of the electron velocity $V_x(t)$ (right column). The number of particles is fixed to $N=10^6$, the time step is $\Delta t=0.0025\,\mathrm{ps}$, the Fermi energy is $\epsilon_F=0.4\,\mathrm{eV}$. The electric field $E_x$ and the GNR width $W$ are varied according to titles of the subfigures.}}
\label{fig:confronto}
\end{figure}

In Figure \ref{fig:confronto}, we compare the solution computed with Algorithm \ref{alg:collision:det} and its Non Parallel version for $N=10^6$. In particular, the time evolution of the energy $\mathcal{E}(t)$ and the longitudinal component of the electron velocity $V_x(t)$, over the time span from $0$ to $3$ ps, are displayed. The parameters are set as in Table \ref{tab:confronto6}, that is $\epsilon_F=0.4$ eV, $E_x=0.5,\,1$ V/$\mu$m, and $W=4,\,7$ nm. We note a good agreement between the two solutions.

Moreover, we compare the CPU time of the Parallel and Non Parallel collision process, for fixed parameters. We choose $N=10^5,\,5\times10^5,\,10^6$ particles, $W=7$ nm, $E_x=0.5$ V/$\mu$m, $\epsilon_F=0.4$ eV, and final time $t=3$ ps with $\Delta t=0.0025$ ps. We store\footnote{Code written in MATLAB R2023a, with an Intel Core i9-12900F 12th Gen processor, 2.40 GHz, and 64,0 GB of RAM.} the time $T^{\mathrm{tot}}$ to compute the time evolution with the different approaches, ignoring the initialization and the post processing, which are independent from the collisional process. In Table \ref{tab:CPUtime} we present the CPU times $T^{\mathrm{tot}}_{NP}$, $T^{\mathrm{tot}}_{P}$, and the ratio $T^{\mathrm{tot}}_{NP}/T^{\mathrm{tot}}_{P}$ to compute the speed-up. We observe that the speed-up obtained with the proposed Algorithm \ref{alg:collision:det} is not negligible.

In what follows, we will therefore restrict ourselves to cases with Fermi energy $\epsilon_F \leq 0.4$~eV and values of the electric field and GNR width as in Figure \ref{fig:confronto}.

\begin{table}[H]
\centering
\begin{footnotesize}
\begin{tabular}{ c|p{3.7cm}|p{3.7cm}|p{3.7cm} }
	\hline
	\multicolumn{4}{c}{$\epsilon_F=0.4$ eV} \\
	\hline
	\rule{0pt}{1\normalbaselineskip}
	& $E_x=0.1$ V/$\mu$m & $E_x=0.5$ V/$\mu$m & $E_x=1$ V/$\mu$m \\
	\hline
	\rule{0pt}{1\normalbaselineskip}
	$W=4$ nm    & $\textrm{E}_\mathcal{E}=0.01\%$, $\textrm{E}_{V_x}=3\%$, \newline $\bar{N}=0$ & $\textrm{E}_\mathcal{E}=0.03\%$, $\textrm{E}_{V_x}=1\%$, \newline $\bar{N}=0$       & $\textrm{E}_\mathcal{E}=0.05\%$, $\textrm{E}_{V_x}=1\%$, \newline $\bar{N}=0$  \\
	\hline
	\rule{0pt}{1\normalbaselineskip}
	$W=7$ nm   & $\textrm{E}_\mathcal{E}=0.03\%$, $\textrm{E}_{V_x}=1.2\%$, \newline $\bar{N}=0$      & $\textrm{E}_\mathcal{E}=0.05\%$, $\textrm{E}_{V_x}=0.6\%$, \newline $\bar{N}=0$       & $\textrm{E}_\mathcal{E}=0.09\%$, $\textrm{E}_{V_x}=0.6\%$, \newline $\bar{N}=0$  \\
	\hline
	\rule{0pt}{1\normalbaselineskip}
	$W=10$ nm   & $\textrm{E}_\mathcal{E}=0.04\%$, $\textrm{E}_{V_x}=1.1\%$, \newline $\bar{N}=0.001\%$ & $\textrm{E}_\mathcal{E}=0.07\%$, $\textrm{E}_{V_x}=0.6\%$, \newline $\bar{N}=0.001\%$ & $\textrm{E}_\mathcal{E}=0.12\%$, $\textrm{E}_{V_x}=0.7\%$, \newline $\bar{N}=0$  \\
	\hline
\end{tabular}
\vspace{0.5cm}

\begin{tabular}{ c|p{3.7cm}|p{3.7cm}|p{3.7cm} }
	\hline
	\multicolumn{4}{c}{$\epsilon_F=0.5$ eV} \\
	\hline
	\rule{0pt}{1\normalbaselineskip}
	& $E_x=0.1$ V/$\mu$m & $E_x=0.5$ V/$\mu$m & $E_x=1$ V/$\mu$m \\
	\hline
	\rule{0pt}{1\normalbaselineskip}
	$W=4$ nm    & $\textrm{E}_\mathcal{E}=0.02\%$, $\textrm{E}_{V_x}=3\%$, \newline $\bar{N}=0$       & $\textrm{E}_\mathcal{E}=0.02\%$, $\textrm{E}_{V_x}=0.9\%$, \newline $\bar{N}=0$       & $\textrm{E}_\mathcal{E}=0.05\%$, $\textrm{E}_{V_x}=0.8\%$, \newline $\bar{N}=0$  \\
		\hline
	\rule{0pt}{1\normalbaselineskip}
	$W=7$ nm    & $\textrm{E}_\mathcal{E}=0.03\%$, $\textrm{E}_{V_x}=1.8\%$, \newline $\bar{N}=0.003\%$ & $\textrm{E}_\mathcal{E}=0.08\%$, $\textrm{E}_{V_x}=0.7\%$, \newline $\bar{N}=0.001\%$ & $\textrm{E}_\mathcal{E}=0.07\%$, $\textrm{E}_{V_x}=0.8\%$, \newline $\bar{N}=0$  \\
		\hline
	\rule{0pt}{1\normalbaselineskip}
	$W=10$ nm   & $\textrm{E}_\mathcal{E}=0.03\%$, $\textrm{E}_{V_x}=1.6\%$, \newline $\bar{N}=0.004\%$ & $\textrm{E}_\mathcal{E}=0.10\%$, $\textrm{E}_{V_x}=0.8\%$, \newline $\bar{N}=0.003\%$ & $\textrm{E}_\mathcal{E}=0.18\%$, $\textrm{E}_{V_x}=1\%$, \newline $\bar{N}=0$  \\
	\hline
\end{tabular}
\vspace{0.5cm}

\begin{tabular}{ c|p{3.7cm}|p{3.7cm}|p{3.7cm} }
	\hline
	\multicolumn{4}{c}{$\epsilon_F=0.6$ eV} \\
	\hline
	\rule{0pt}{1\normalbaselineskip}
	& $E_x=0.1$ V/$\mu$m & $E_x=0.5$ V/$\mu$m & $E_x=1$ V/$\mu$m \\
	\hline
	\rule{0pt}{1\normalbaselineskip}
	$W=4$ nm   & $\textrm{E}_\mathcal{E}=0.02\%$, $\textrm{E}_{V_x}=3.5\%$, \newline $\bar{N}=0$       & $\textrm{E}_\mathcal{E}=0.03\%$, $\textrm{E}_{V_x}=1.1\%$, \newline $\bar{N}=0$       & $\textrm{E}_\mathcal{E}=0.05\%$, $\textrm{E}_{V_x}=0.9\%$, \newline $\bar{N}=0$        \\
	\hline
	\rule{0pt}{1\normalbaselineskip}
	$W=7$ nm   & $\textrm{E}_\mathcal{E}=0.03\%$, $\textrm{E}_{V_x}=2.7\%$, \newline $\bar{N}=0.009\%$ & $\textrm{E}_\mathcal{E}=0.05\%$, $\textrm{E}_{V_x}=0.8\%$, \newline $\bar{N}=0.005\%$ & $\textrm{E}_\mathcal{E}=0.16\%$, $\textrm{E}_{V_x}=0.9\%$, \newline $\bar{N}=0.002\%$  \\
	\hline
	\rule{0pt}{1\normalbaselineskip}
	$W=10$ nm   & $\textrm{E}_\mathcal{E}=0.03\%$, $\textrm{E}_{V_x}=2\%$, \newline $\bar{N}=0.007\%$ & $\textrm{E}_\mathcal{E}=0.07\%$, $\textrm{E}_{V_x}=1.1\%$, \newline $\bar{N}=0.004\%$ & $\textrm{E}_\mathcal{E}=0.15\%$, $\textrm{E}_{V_x}=1\%$, \newline $\bar{N}=0.001\%$  \\
	\hline
\end{tabular}
\caption{\small{Relative error between the mean energy and velocity curves obtained with the P and the NP algorithms in the case of $N=10^5$ particles. The percentage of particles that violate the Pauli principle is also reported.}}
\label{tab:confronto5}
\end{footnotesize}
\end{table}
\begin{table}[H] 
	\centering
\begin{footnotesize}
	\begin{tabular}{ c|p{3.7cm}|p{3.7cm}|p{3.7cm} }
		\hline
		\multicolumn{4}{c}{$\epsilon_F=0.4$ eV} \\
		\hline
		\rule{0pt}{1\normalbaselineskip}
	& $E_x=0.1$ V/$\mu$m & $E_x=0.5$ V/$\mu$m & $E_x=1$ V/$\mu$m \\
		\hline
		\rule{0pt}{1\normalbaselineskip}
		$W=4$ nm    & $\mathrm{E}_\mathcal{E}=0.01\%$, $\mathrm{E}_{V_x}=1\%$, \newline $\bar{N}=0$        & $\mathrm{E}_\mathcal{E}=0.01\%$, $\mathrm{E}_{V_x}=0.5\%$, \newline $\bar{N}=0$       & $\mathrm{E}_\mathcal{E}=0.02\%$, $\mathrm{E}_{V_x}=0.6\%$, \newline $\bar{N}=0$   \\
		\hline
		\rule{0pt}{1\normalbaselineskip}
		$W=7$ nm    & $\mathrm{E}_\mathcal{E}=0.01\%$, $\mathrm{E}_{V_x}=0.4\%$, \newline $\bar{N}=0$        & $\mathrm{E}_\mathcal{E}=0.05\%$, $\mathrm{E}_{V_x}=0.3\%$, \newline $\bar{N}=0$       & $\mathrm{E}_\mathcal{E}=0.08\%$, $\mathrm{E}_{V_x}=0.4\%$, \newline $\bar{N}=0$   \\
		\hline
		\rule{0pt}{1\normalbaselineskip}
		$W=10$ nm   & $\mathrm{E}_\mathcal{E}=0.02\%$, $\mathrm{E}_{V_x}=0.4\%$, \newline $\bar{N}=0$        & $\mathrm{E}_\mathcal{E}=0.07\%$, $\mathrm{E}_{V_x}=0.4\%$, \newline $\bar{N}=0$       & $\mathrm{E}_\mathcal{E}=0.13\%$, $\mathrm{E}_{V_x}=0.4\%$, \newline $\bar{N}=0$   \\
		\hline
	\end{tabular}
	\vspace{0.5cm}
	
	\begin{tabular}{ c|p{3.7cm}|p{3.7cm}|p{3.7cm} }
		\hline
		\multicolumn{4}{c}{$\epsilon_F=0.5$ eV} \\
		\hline
		\rule{0pt}{1\normalbaselineskip}
	& $E_x=0.1$ V/$\mu$m & $E_x=0.5$ V/$\mu$m & $E_x=1$ V/$\mu$m \\
		\hline
		\rule{0pt}{1\normalbaselineskip}
		$W=4$ nm    & $\mathrm{E}_\mathcal{E}=0.01\%$, $\mathrm{E}_{V_x}=1\%$, \newline $\bar{N}=0$        & $\mathrm{E}_\mathcal{E}=0.01\%$, $\mathrm{E}_{V_x}=0.5\%$, \newline $\bar{N}=0$       & $\mathrm{E}_\mathcal{E}=0.02\%$, $\mathrm{E}_{V_x}=0.4\%$, \newline $\bar{N}=0$         \\
		\hline
		\rule{0pt}{1\normalbaselineskip}
		$W=7$ nm   & $\mathrm{E}_\mathcal{E}=0.01\%$, $\mathrm{E}_{V_x}=0.8\%$, \newline $\bar{N}=0.0001\%$ & $\mathrm{E}_\mathcal{E}=0.07\%$, $\mathrm{E}_{V_x}=0.5\%$, \newline $\bar{N}=0$       & $\mathrm{E}_\mathcal{E}=0.09\%$, $\mathrm{E}_{V_x}=0.6\%$, \newline $\bar{N}=0$         \\
		\hline
		\rule{0pt}{1\normalbaselineskip}
		$W=10$ nm  & $\mathrm{E}_\mathcal{E}=0.02\%$, $\mathrm{E}_{V_x}=0.7\%$, \newline $\bar{N}=0.0003\%$ & $\mathrm{E}_\mathcal{E}=0.07\%$, $\mathrm{E}_{V_x}=0.4\%$, \newline $\bar{N}=0$       & $\mathrm{E}_\mathcal{E}=0.14\%$, $\mathrm{E}_{V_x}=0.5\%$, \newline $\bar{N}=0.0001\%$  \\
		\hline
	\end{tabular}
	\vspace{0.5cm}
	
	\begin{tabular}{ c|p{3.7cm}|p{3.7cm}|p{3.7cm} }
		\hline
		\multicolumn{4}{c}{$\epsilon_F=0.6$ eV} \\
		\hline
		\rule{0pt}{1\normalbaselineskip}
	& $E_x=0.1$ V/$\mu$m & $E_x=0.5$ V/$\mu$m & $E_x=1$ V/$\mu$m \\
		\hline
		\rule{0pt}{1\normalbaselineskip}
		$W=4$ nm   & $\mathrm{E}_\mathcal{E}=0.01\%$, $\mathrm{E}_{V_x}=1.2\%$, \newline $\bar{N}=0$        & $\mathrm{E}_\mathcal{E}=0.03\%$, $\mathrm{E}_{V_x}=0.6\%$, \newline $\bar{N}=0$        & $\mathrm{E}_\mathcal{E}=0.05\%$, $\mathrm{E}_{V_x}=0.5\%$, \newline $\bar{N}=0$         \\
		\hline
		\rule{0pt}{1\normalbaselineskip}
		$W=7$ nm    & $\mathrm{E}_\mathcal{E}=0.01\%$, $\mathrm{E}_{V_x}=0.9\%$, \newline $\bar{N}=0.0004\%$ & $\mathrm{E}_\mathcal{E}=0.05\%$, $\mathrm{E}_{V_x}=0.6\%$, \newline $\bar{N}=0.0001\%$ & $\mathrm{E}_\mathcal{E}=0.11\%$, $\mathrm{E}_{V_x}=0.5\%$, \newline $\bar{N}=0.0002\%$  \\
		\hline
		\rule{0pt}{1\normalbaselineskip}
		$W=10$ nm  & $\mathrm{E}_\mathcal{E}=0.01\%$, $\mathrm{E}_{V_x}=0.8\%$, \newline $\bar{N}=0.0010\%$ & $\mathrm{E}_\mathcal{E}=0.07\%$, $\mathrm{E}_{V_x}=0.5\%$, \newline $\bar{N}=0.0003\%$ & $\mathrm{E}_\mathcal{E}=0.14\%$, $\mathrm{E}_{V_x}=0.6\%$, \newline $\bar{N}=0$         \\
		\hline
	\end{tabular}
	\caption{\small{Relative error between the mean energy and velocity  obtained with the P and the NP algorithms in the case of $N=10^6$ particles. The percentage of particles that violate the Pauli principle is also reported.}}
	\label{tab:confronto6}
\end{footnotesize}
\end{table}
\begin{table}[H]
\centering
\begin{footnotesize}
\begin{tabular}{ c c c c c c c }
	\hline
	\rule{0pt}{1\normalbaselineskip}
	                 && $N=10^5$ && $N=5\times10^5$ && $N=10^6$ \\[2pt]
	\hline
	\rule{0pt}{1\normalbaselineskip}
	$T^{\mathrm{tot}}_{NP}$ && $825.86$ s      && $4.12\times10^3$ s   && $8.03\times10^3$ s  \\[4pt]
	\rule{0pt}{1\normalbaselineskip}
	$T^{\mathrm{tot}}_{P}$  && $147.75$ s      && $0.78\times10^3$ s   && $1.60\times10^3$ s  \\[4pt]
	\rule{0pt}{1\normalbaselineskip}
	$\frac{T^{\mathrm{tot}}_{NP}}{T^{\mathrm{tot}}_{P}}$ && $\sim5.6$   && $\sim5.3$   && $\sim5.0$  \\[6pt]
	\hline	
\end{tabular}
\caption{\small{Comparison of CPU time of the Parallel and Non Parallel collision process. We fix $W=7$ nm, $E_x=0.5$ V/$\mu$m, $\epsilon_F=0.4$ eV, and final time $t=3$ ps with $\Delta t=0.0025$ ps. We vary the number of particles $N=10^5,\,5\times10^5,\,10^6$. The last line represents the approximated speed-up.}}
\label{tab:CPUtime}
\end{footnotesize}
\end{table}
\subsection{Test 2: Semiclassical BE in the presence of uncertainties}
In this subsection, the semiclassical Boltzmann equation in the presence of uncertainties is considered. In particular, we adopt Algorithm \ref{alg:collision:unc} and investigate two different scenarios. In the first one, we take into account an uncertainty in the GNR width
\[
W = (6.5 + \z) \, \mbox{nm}, \qquad \textrm{with} \qquad \z\sim\mathcal{U}([0,1])
\]
being $\mathcal{U}(\cdot)$ the uniform distribution, and consider a deterministic electric field $E_x=0.5,\,1$ V/$\mu$m, and several values for the Fermi energy $\epsilon_F=0.2,\,0.4$ eV. In the second one, the uncertainty is introduced in the electric field
\[
E_x = (0.4 + 0.2\z) \, \mbox{V/$\mu$m}, \qquad \textrm{with} \qquad \z\sim\mathcal{U}([0,1]),
\]
while $W=4,\,7$ nm and $\epsilon_F=0.2,\,0.4$ eV. In all simulations we set $N=10^6$ particles and $M=5$ as gPC truncation order.

\begin{figure}
	\centering
	\includegraphics[width = 0.3\linewidth]{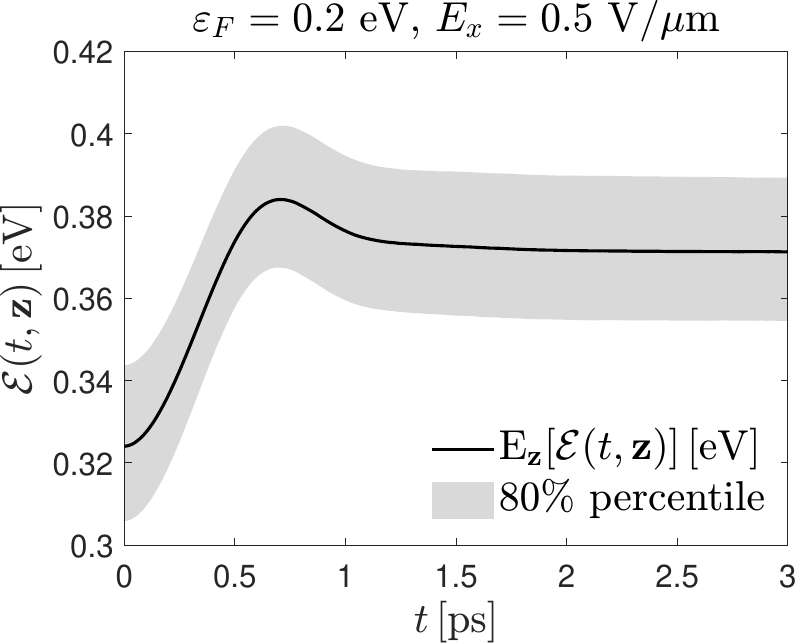}\hspace{2ex}
	\includegraphics[width = 0.3\linewidth]{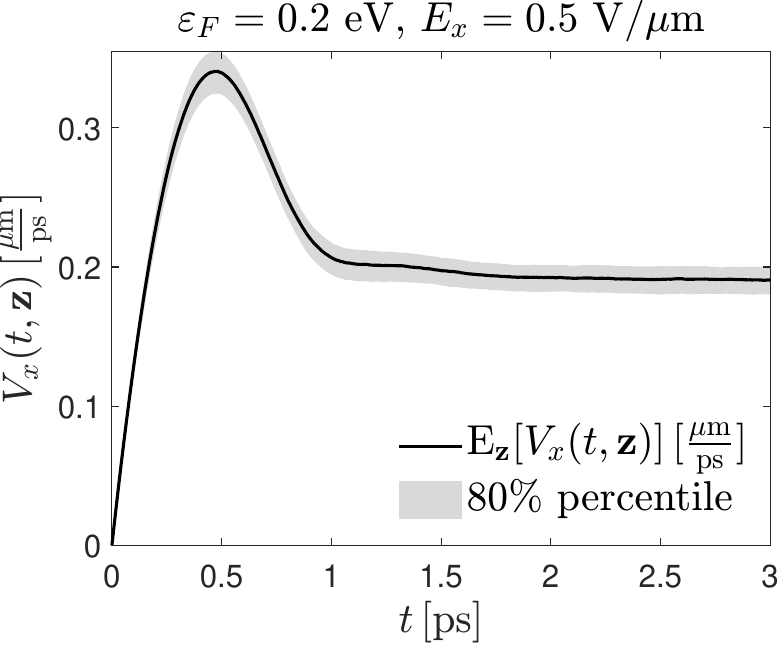} \\
	\vspace{2ex}
	
	\includegraphics[width = 0.3\linewidth]{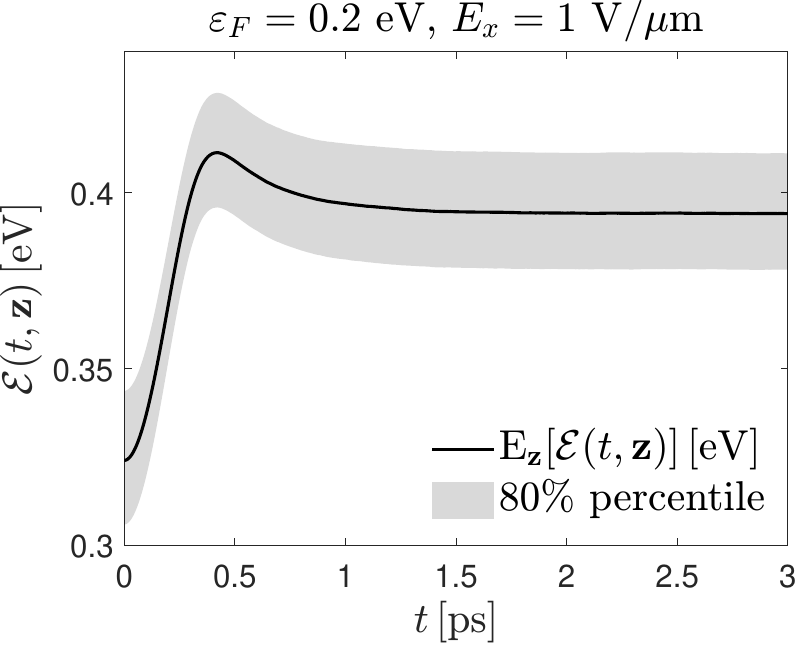}\hspace{2ex}
	\includegraphics[width = 0.3\linewidth]{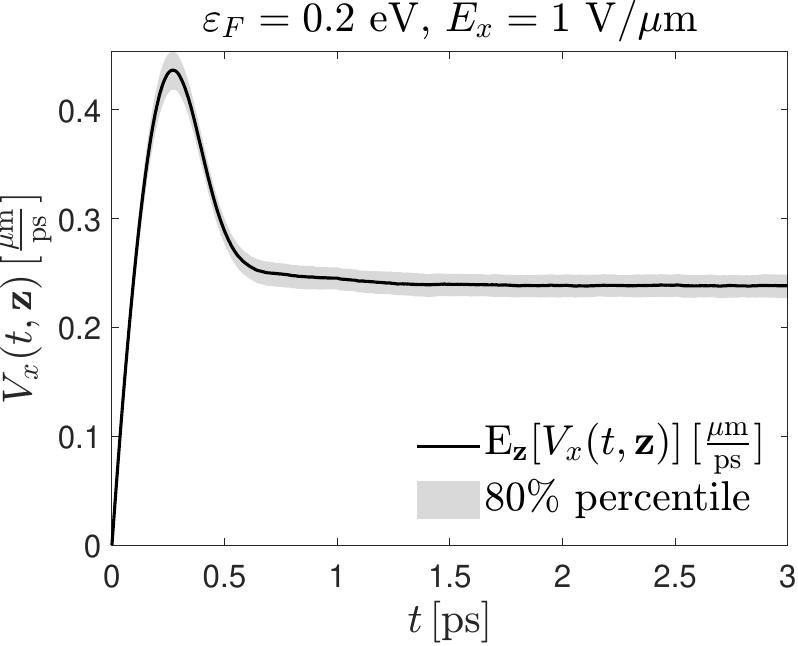} \\
	\vspace{2ex}
		
	\includegraphics[width = 0.3\linewidth]{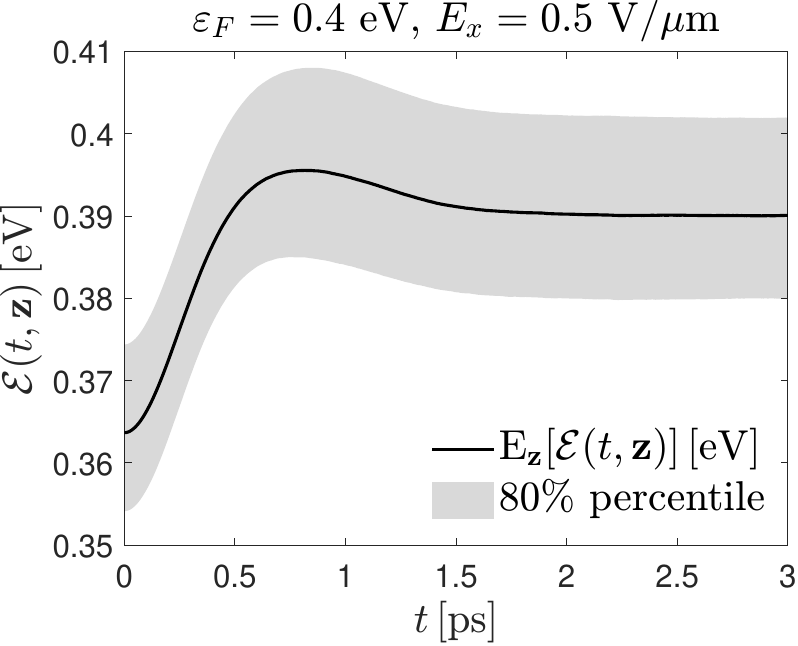}\hspace{2ex}
	\includegraphics[width = 0.3\linewidth]{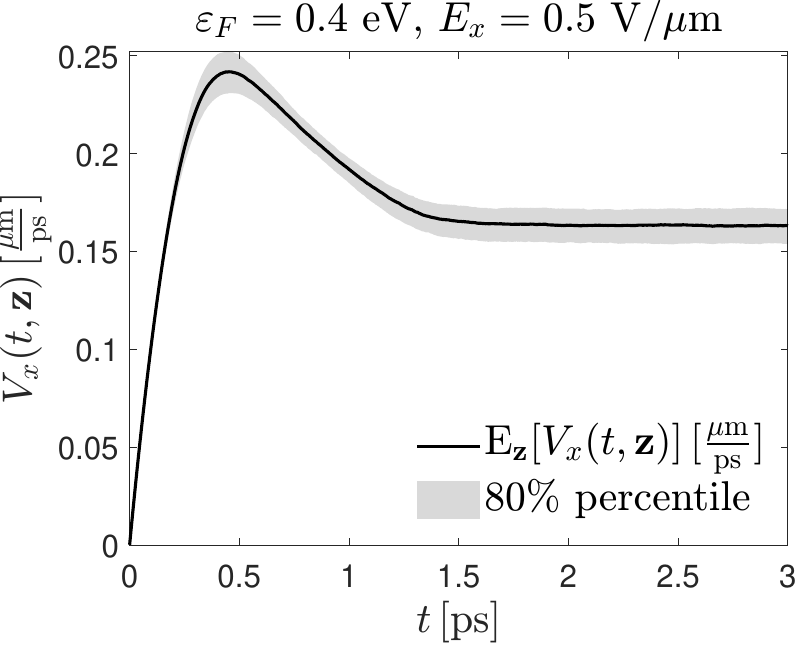} \\
	\vspace{2ex}	
	
	\includegraphics[width = 0.3\linewidth]{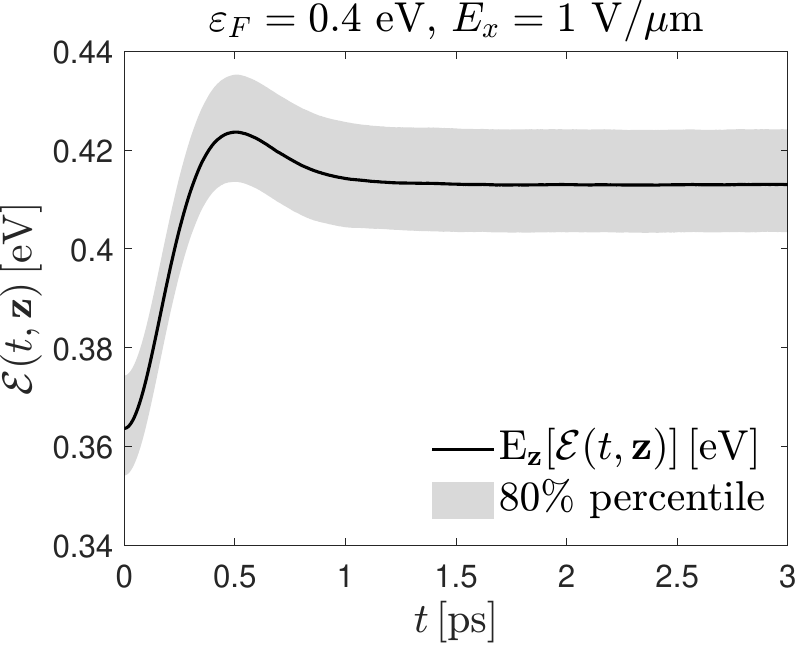}\hspace{2ex}
	\includegraphics[width = 0.3\linewidth]{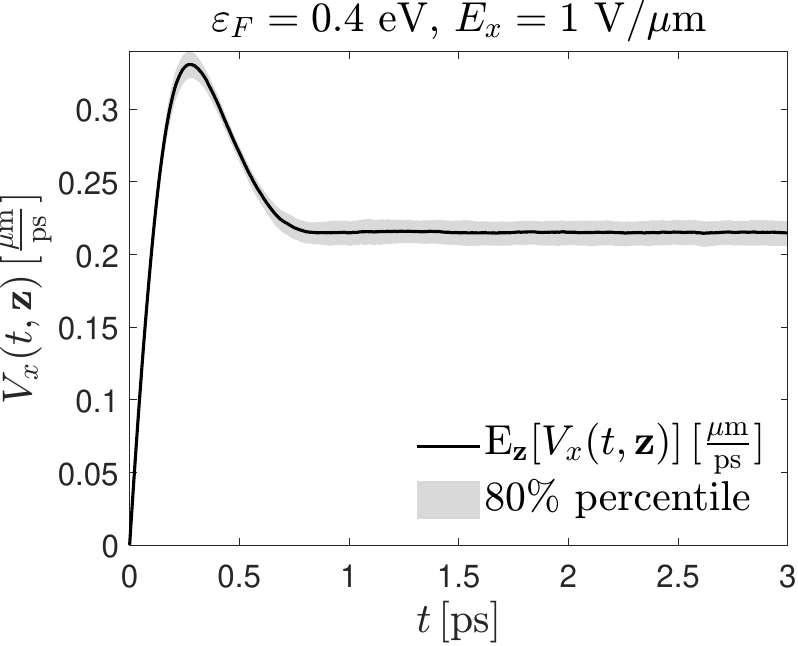} 
	\caption{\small{\textbf{Test 2}: Time evolution of the expectations w.r.t. $\z$ of the energy $\mathcal{E}(t,\z)$ (left column) and the longitudinal component of the electron velocity $V_x(t,\z)$ (right column). In all the simulations, we consider uncertainties in the GNR width $W(\z)=(6.5+\z) \, \mathrm{nm}$, with $\z\sim\mathcal{U}([0,1])$. The number of particles is $N=10^6$, the time step is $\Delta t = 0.0025 \, \mathrm{ps}$. The Fermi energy $\epsilon_F$ and the electric field $E_x$ are varied according to the titles of the subfigures.}}
	\label{fig:Wz}
\end{figure}

In Figure \ref{fig:Wz}  the results with the uncertainty in the width of the GNR are plotted along with the 80th percentile for the electron energy and velocity. The corresponding confidence intervals are close to the mean values in all the simulated cases revealing a robust behaviour of the GNR with respect to physically reasonable variations in the GNR width. The deviation from the mean value is less than 10\% for both energy and velocity. 

\begin{figure}
	\centering
	\includegraphics[width = 0.3\linewidth]{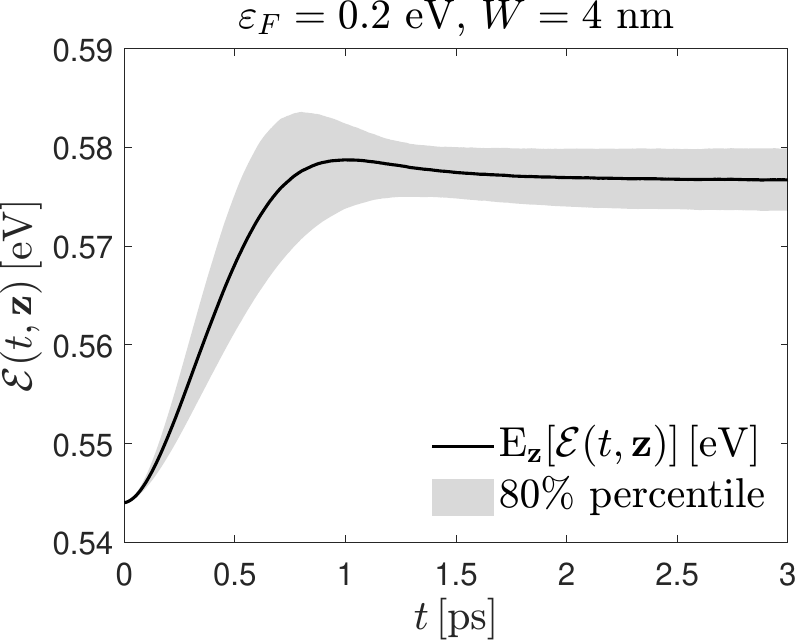}\hspace{2ex}
	\includegraphics[width = 0.3\linewidth]{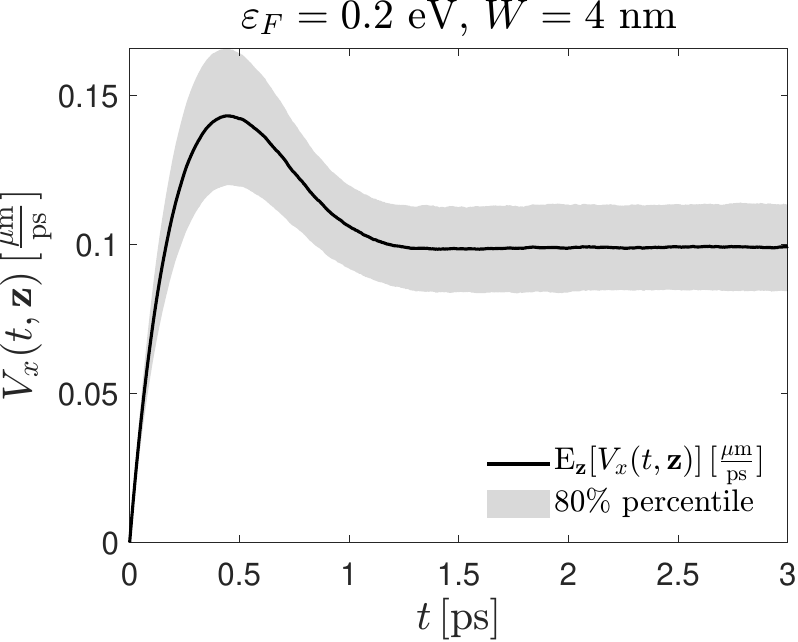} \\
	\vspace{2ex}
	
	\includegraphics[width = 0.3\linewidth]{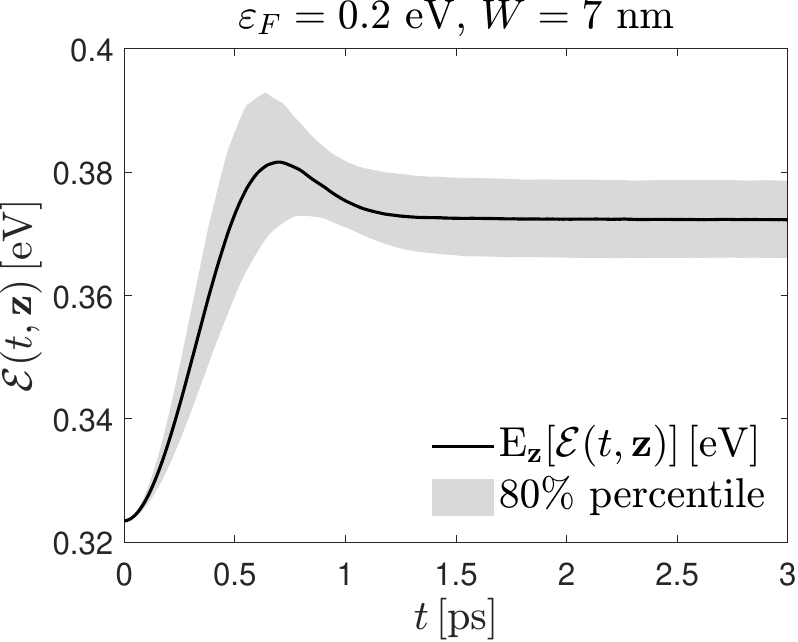}\hspace{2ex}
	\includegraphics[width = 0.3\linewidth]{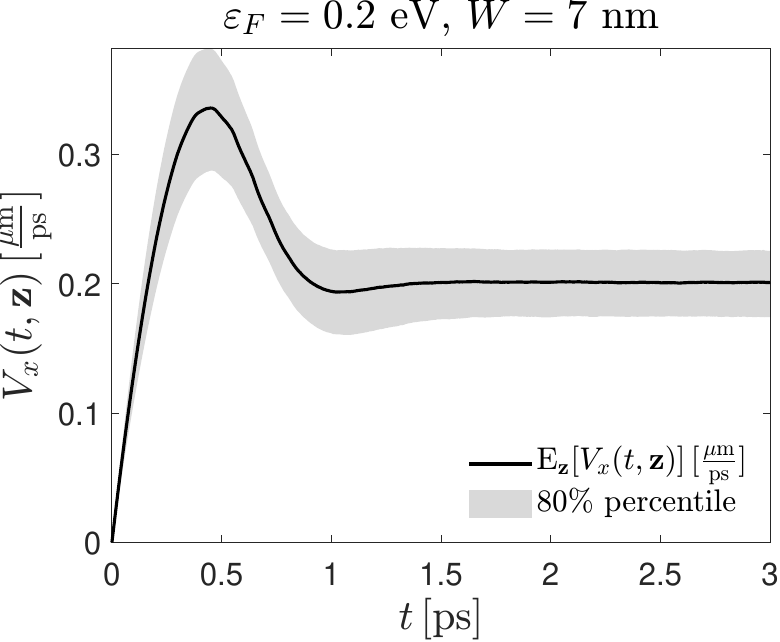} \\
	\vspace{2ex}
	
	\includegraphics[width = 0.3\linewidth]{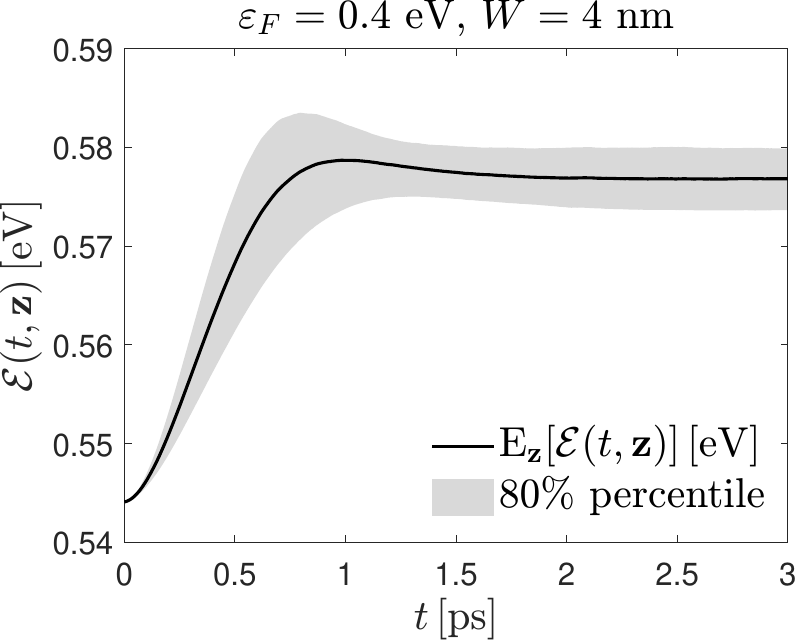}\hspace{2ex} 
	\includegraphics[width = 0.3\linewidth]{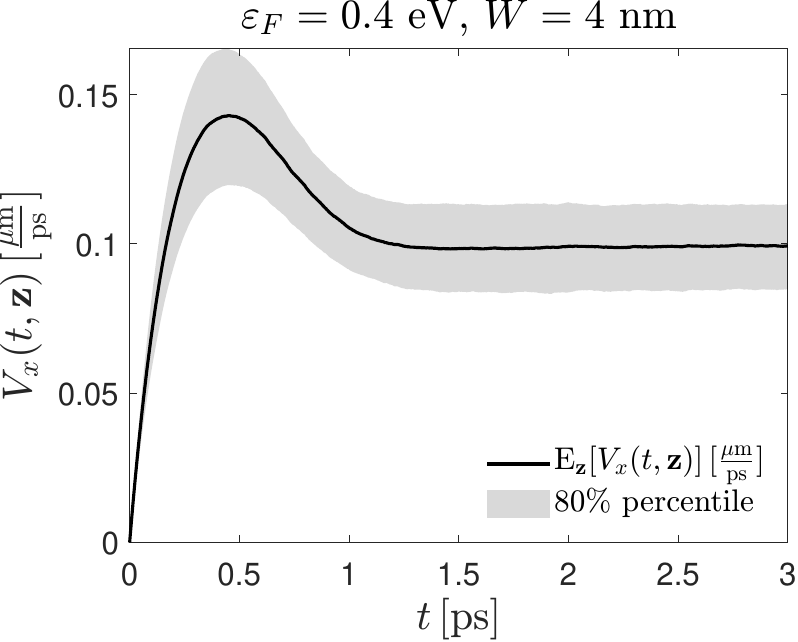} \\
	\vspace{2ex}	
	
	\includegraphics[width = 0.3\linewidth]{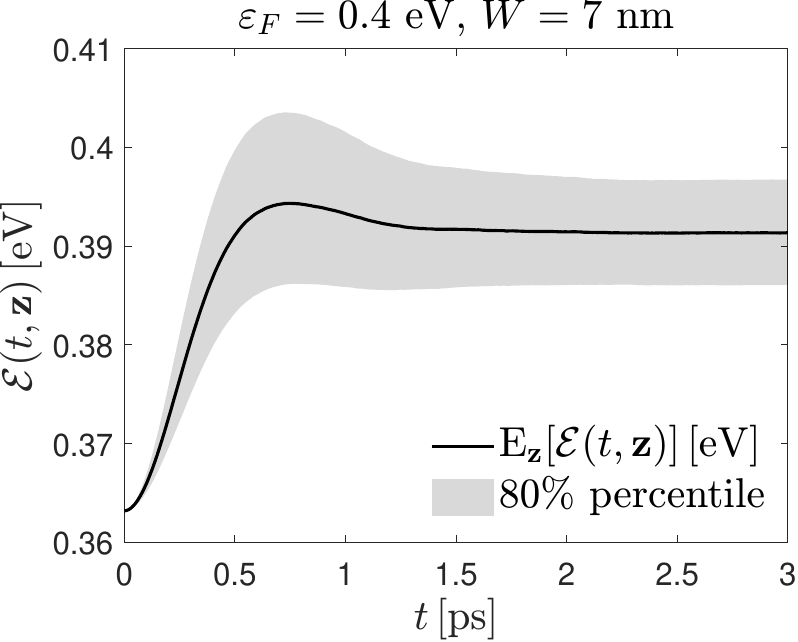}\hspace{2ex}
	\includegraphics[width = 0.3\linewidth]{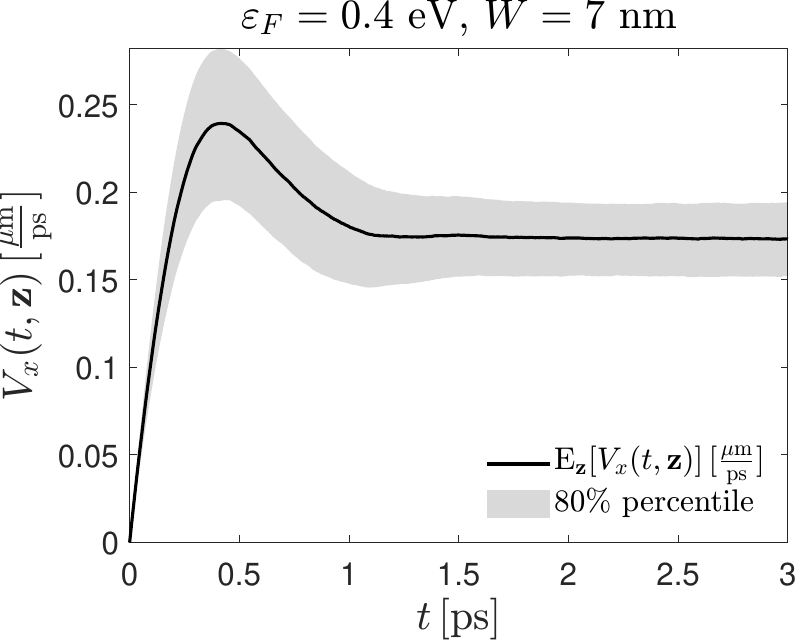} 
	\caption{\small{\textbf{Test 2}:  Time evolution of the expectations w.r.t. $\z$ of the energy $\mathcal{E}(t,\z)$ (left column) and the longitudinal component of the electron velocity $V_x(t,\z)$ (right column). In all the simulations, we consider uncertainties in the applied electric field $E_x(\z)=(0.4+0.2\z) \, \mathrm{V} / \mu \mathrm{m}$, with $\z\sim\mathcal{U}([0,1])$. The number of particles is $N=10^6$, the time step is $\Delta t = 0.0025 \, \mathrm{ps}$. The Fermi energy $\epsilon_F$ and the GNR width $W$ are varied according to the titles of the subfigures.}}
	\label{fig:Ez}
\end{figure}

In Figure \ref{fig:Ez}  the results with the uncertainty in the electric field  are plotted along with the 80\% percentile for the electron energy and the velocity. The corresponding confidence intervals are a bit wider than the previous case showing a more accentuated sensitivity of the behaviour on the applied field. However,  the confidence interval is still acceptable to  guarantee again a robust behaviour of the GNR against physically reasonable variations in the applied electric field.

\subsection{Test 3: Stochastic Galerkin error}
\begin{figure}
	\centering
	\includegraphics[width = 0.45\linewidth]{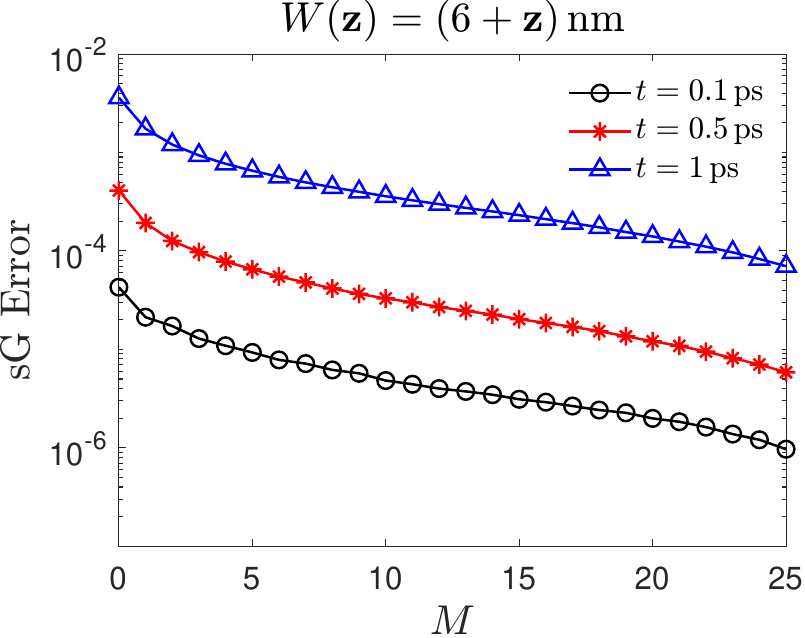}\hspace{2ex}
	\includegraphics[width = 0.45\linewidth]{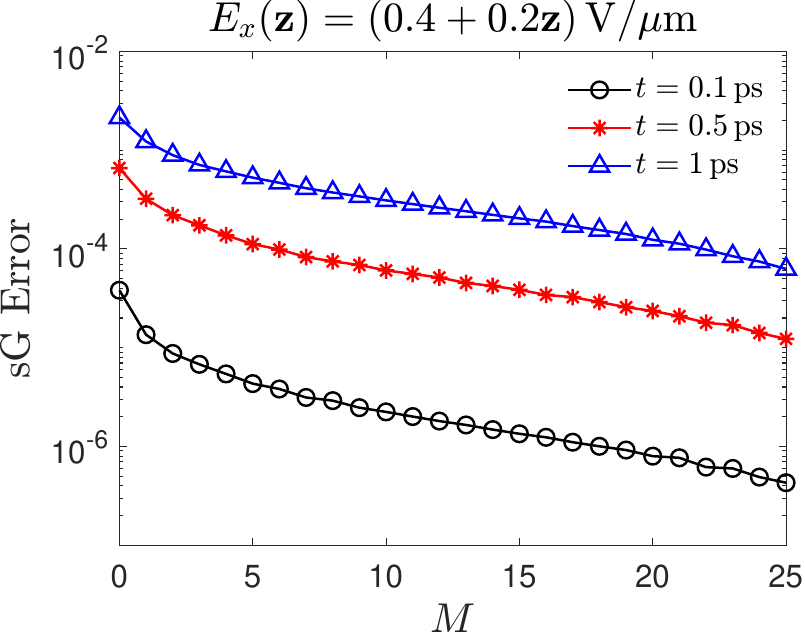}
	\caption{\small{\textbf{Test 3}: Comparison of the sG Error in the evaluation of the energy $\mathcal{E}(t,\z)$ at fixed times $t=0.1,\,0.5,\,1\,\mathrm{ps}$, for increasing $M$. Left panel: uncertain GNR width $W(\z) = ( 6.5 + \z ) \, \textrm{nm}$, $\z\sim\mathcal{U}([0,1])$ and $E_x=0.5\,\mathrm{V}/\mu\mathrm{m}$. Right panel: uncertain electric field $E_x(\z) = (0.4+0.2\z) \, \mathrm{V}/\mu\mathrm{m}$, $\z\sim\mathcal{U}([0,1])$, and $W=7\, \textrm{nm}$. In all the simulations $N=10^4$, the time step is $\Delta t=0.0025\,\mathrm{ps}$, and the Fermi level is $\epsilon_F=0.4\,\mathrm{eV}$. Reference solution computed with $M^{\textrm{ref}}=30$. }}
	\label{fig:sGerror}
\end{figure}

In this section, we investigate numerically the convergence in the space of the random parameters of the solution obtained with Algorithm \ref{alg:collision:unc}. In particular, we compute the $L^2$ error at fixed times $t=0.1,\,0.5,\,1\,\mathrm{ps}$ in the evaluation of the energy
\[
\textrm{sG Error} = \| \mathcal{E}^{\textrm{ref}}(t,\z) - \mathcal{E}(t,\z)\|_{L^2(I_\z)},
\]
with respect to a reference solution with $M^{\textrm{ref}}=30$, for increasing $M=0,\dots,25$. We store the initial data and the collisional tree, to evaluate the error in the polynomial approximation of the particles. In Figure \ref{fig:sGerror} we consider two scenarios: the left panel corresponds to an uncertain GNR width 
\[
W(\z) = ( 6.5 + \z ) \, \textrm{nm},\qquad \z\sim\mathcal{U}([0,1])
\]
and a deterministic applied electric field $E_x=0.5\,\mathrm{V}/\mu\mathrm{m}$; the right panel to an uncertain electric field
\[
E_x(\z) = (0.4+0.2\z) \, \mathrm{V}/\mu\mathrm{m},\qquad \z\sim\mathcal{U}([0,1])
\]
and a deterministic GNR width $W=7\, \textrm{nm}$. In both cases the number of particles is fixed to $N=10^4$, the time step is $\Delta t=0.0025\,\mathrm{ps}$, and the Fermi level is $\epsilon_F=0.4\,\mathrm{eV}$. 

\rev{As we may notice looking at Figure \ref{fig:sGerror}, the spectral accuracy deteriorates as the time increases, although the decreasing rates seem approximately the same. This is due to the presence of indicator functions in the collisional process: the more collisions occur, the higher the sG error is. Despite this, in the relevant time span on the evolution of the observables $\mathcal{E}(t,\z)$ and $V_x(t,\z)$ (look, e.g., at Figure \ref{fig:Wz}-\ref{fig:Ez}), with truncation order $M=5$ we observe a good accuracy.}

\subsection{Test 4: Comparison with Monte Carlo sampling method}
\begin{figure}
	\centering
	\includegraphics[width = 0.45\linewidth]{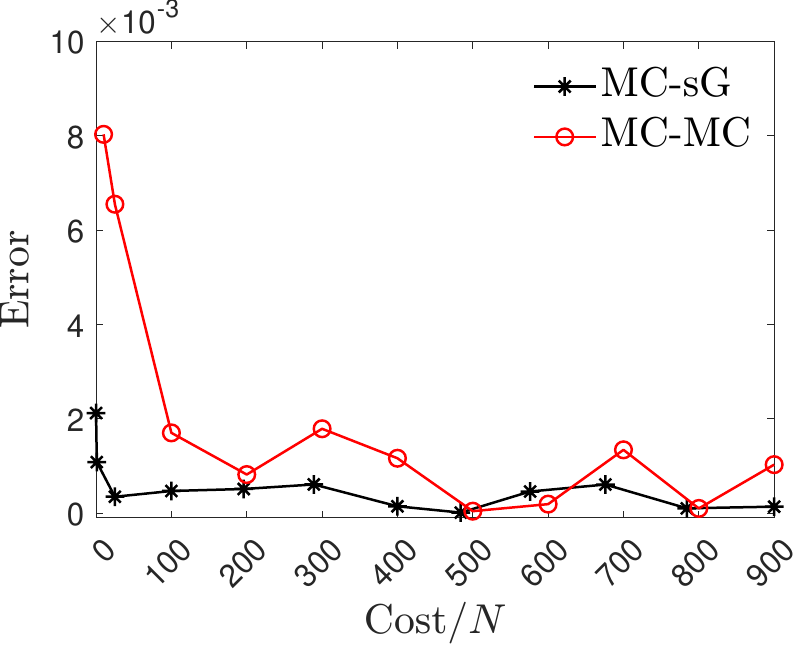}
	\caption{\small{\textbf{Test 4}: Comparison Error-Cost/$N$ between the particle stochastic Galerkin scheme (MC-sG) and a Monte Carlo sampling of the random parameter (MC-MC). Error computed as \eqref{eq:errL1} at the fixed time $t=0.5$ ps, with $N=5\times10^4$. Reference solution computed with $5\times10^6$ particles and $M^{\textrm{ref}}=30$.}}
	\label{fig:sGMC}
\end{figure}

The aim of this section is to compare the method proposed in Algorithm \ref{alg:collision:unc}, referred to as MC-sG, with deterministic Algorithm \ref{alg:collision:det} together with a standard Monte Carlo sampling of the random parameter, which we indicate as MC-MC. The cost of MC-sG is $\rev{O(N\cdot M_1^2)}$, where $N$ is the number of particles and \rev{$M_1$} the order of truncation of the gPC expansion. On the other hand, the cost of MC-MC is $\rev{O(N\cdot M_2)}$, \rev{where $M_2$} is the number of random samples of $\z$. First, we compute a reference solution up to time $t=0.5$ ps with the particle stochastic Galerkin method by using $5\times10^6$ particles and an order of expansion $M^{\textrm{ref}}=30$. Then, we fix $N=5\times10^4$ and we compute for both MC-sG and MC-MC the error in the evaluation of $V_x(t,\z)$ against the reference solution
\begin{equation} \label{eq:errL1}
\textrm{Error}= \left| \E_{\z}[V_x(t,\z)] - \E_{\z}[V^{\textrm{ref}}_x(t,\z)] \right|.
\end{equation}
In Figure \ref{fig:sGMC} we display the error as a function of the computational cost divided by the number of particles $N$, \rev{which is $M_1^2$ for MC-sG and $M_2$ for MC-MC}. In can be noticed that for small costs the MC-sG Algorithm \ref{alg:collision:unc} performs better than the MC-MC, while for higher costs the errors obtained with the two methods are comparable and there are no substantial differences.

\section*{Conclusions} 
An efficient method to include the uncertainty of relevant  parameters in phenomena described by a Boltzmann transport equation has been devised for the analysis of graphene nanoribbons. The proposed scheme is substantially in agreement with the Pauli exclusion principle and includes generalized \rev{Polynomial Chaos} also in the presence of complex scattering terms in the numerical scheme for the Boltzmann equation based on a discontinuous Galerkin method.  The quantification of the uncertainty in the average electron velocity and energy has been established showing a good robustness in the electrical performance of GNRs.

\rev{Future research directions will be devoted to the study of adaptive-type methods in which the initial distribution of random parameters evolves over time due to the dynamics. This has been done, e.g., in \cite{gerritsma2010}.}

\section*{Acknowledgments} 
The authors acknowledge the support from INdAM (GNFM). 
The author V.R. acknowledges the support from MUR project PRIN 2022 ``Transport phenomena in low dimensional structures: models, simulations and theoretical aspect'' CUP E53D23005900006. 
G.N. acknowledges the financial support from the project PON R\&I 2014-2020 ``Asse IV - Istruzione e ricerca per il recupero - REACT-EU, Azione IV.4 - Dottorati e contratti di ricerca su tematiche dell'innovazione'', project ``Modellizzazione, simulazione e design di transistori innovativi''. 
A.M. and G.N.  acknowledge the support from  Progetto di Ricerca GNFM-INDAM 2023 ``Uncertainty quantification for kinetic models describing physical and socio-economical
phenomena'' CUP E53C2200193000. 
M.Z. acknowledges the support of the ICSC – Centro Nazionale di Ricerca in High Performance Computing, Big Data and Quantum Computing, funded by European Union - NextGeneration EU. M. Z. wishes also to acknowledge partial support by MUR-PRIN2022PNRR Project No. P2022Z7ZAJ.

\appendix
\section{Appendix}\label{App_sc_rate}
We provide the explicit expressions for the scattering rates of all the considered types of phonons, described in Sec. \ref{SEC:BE}. If $\nu\in\left\lbrace AC, OP, K \right\rbrace$ the scattering rates are defined by
\rev{\begin{align*}
	\Gamma^{(\nu)} = & \int_{\mathbb{R}^2} S^{(\nu)}(\bk,\bk') \,d\bk'\\
	= &\int_{\mathbb{R}^2} \left| G^{(\nu)}(\bk, \bk') \right|^{2} \bigg[ \left( n^{(\nu)}_{\mathbf{q}} + 1 \right) \delta \left( \eps(\bk') - \eps(\bk) + \hbar \, \omega^{(\nu)}_{\mathbf{q}} \right) \\
	& + n^{(\nu)}_{\mathbf{q}} \, \delta \left( \eps(\bk') - \eps(\bk) - \hbar \, \omega^{(\nu)}_{\mathbf{q}} \right) \bigg] \, d\bk' .
\end{align*}}
We make the change of variables $(k'_x,k'_y)\rightarrow(\eps',\theta')$ such that
$$
\left\lbrace
\begin{aligned}
	k'_x & = \vert \bk'\vert \cos\theta' = \sqrt{\frac{\eps'^2}{(\hbar v_F)^2}-\left(\frac{\pi}{W}\right)^2}\cos\theta'\\
	k'_y & = \vert \bk'\vert \sin\theta' = \sqrt{\frac{\eps'^2}{(\hbar v_F)^2}-\left(\frac{\pi}{W}\right)^2}\sin\theta'
\end{aligned}
\right.
$$ 
with $(\eps',\theta')\in [\eps_0,+\infty[\times[0,2\pi]$ where $\eps_0=\hbar v_F\frac{\pi}{W}$. The Jacobian of such a transformation is $\dfrac{\eps'}{(\hbar v_F)^2}$. By defining $\bar{G}^{(\nu)}(\theta,\theta')=G^{(\nu)}(\bk,\bk')$ one has
\rev{\begin{align*}
	&\Gamma^{(\nu)} = \frac{1}{(\hbar v_F)^2} \int_0^{2\pi} \left| \tilde{G}^{(\nu)}(\theta,\theta') \right|^{2} \, d\theta' \cdot \int_{\mathbb{R}} \left[ \left( n^{(\nu)}_{\mathbf{q}} + 1 \right) \delta \left( \eps' - \eps + \hbar \, \omega^{(\nu)}_{\mathbf{q}} \right) \right. \\
	& \qquad \left. + n^{(\nu)}_{\mathbf{q}} \, \delta \left( \eps' - \eps - \hbar \, \omega^{(\nu)}_{\mathbf{q}} \right) \right] \chi\left(\eps'\in [\eps_0,+\infty[\right) \eps' \, d\eps'\\
	& = \frac{1}{(\hbar v_F)^2} \int_0^{2\pi} \left| \tilde{G}^{(\nu)}(\theta,\theta') \right|^{2} \, d\theta' \cdot \left[ \left( n^{(\nu)}_{\mathbf{q}} + 1 \right) \left( \eps - \hbar \, \omega^{(\nu)}_{\mathbf{q}} \right) H \left( \eps - \hbar \, \omega^{(\nu)}_{\mathbf{q}} \right) H\left( \eps - \hbar \, \omega^{(\nu)}_{\mathbf{q}} -\eps_0 \right) \right. \\
	& \qquad \left. + n^{(\nu)}_{\mathbf{q}} \, \left( \eps + \hbar \, \omega^{(\nu)}_{\mathbf{q}} \right) H\left( \eps + \hbar \, \omega^{(\nu)}_{\mathbf{q}} -\eps_0 \right) \right] \\
	& = \frac{1}{(\hbar v_F)^2} \int_0^{2\pi} \left| \tilde{G}^{(\nu)}(\theta,\theta') \right|^{2} \, d\theta' \cdot \left[ \left( n^{(\nu)}_{\mathbf{q}} + 1 \right) \left( \eps - \hbar \, \omega^{(\nu)}_{\mathbf{q}} \right) H\left( \eps - \hbar \, \omega^{(\nu)}_{\mathbf{q}} -\eps_0 \right) \right. \\
	& \qquad \left. + n^{(\nu)}_{\mathbf{q}} \, \left( \eps + \hbar \, \omega^{(\nu)}_{\mathbf{q}} \right) H\left( \eps + \hbar \, \omega^{(\nu)}_{\mathbf{q}} -\eps_0 \right) \right] ,
\end{align*}} 
where $H(\cdot)$ represents the Heaviside step function. The expressions for the acoustic case can be formally recovered by  formally taking the limit $\hbar\omega_\mathbf{q}^{(AC)} \to 0$.

Explicitly, if $\nu=ac$ we have
\begin{equation*}
	\Gamma^{(ac)} = \frac{1}{(\hbar v_F)^2} \frac{1}{(2\pi)^2} \frac{\pi D_{ac}^2 k_B T}{2\hbar\sigma_m v_p^2}2\pi \eps = \frac{D_{ac}^2 k_B T}{4\hbar^3 v_F^2 \sigma_m v_p^2}\eps H(\eps-\eps_0),
\end{equation*}
while if $\nu=O$ we get
\begin{align*}
	\Gamma^{(O)} & = \frac{1}{(\hbar v_F)^2} \frac{2}{(2\pi)^2} \frac{\pi D_O^2}{\sigma_m \omega_O} 2\pi \left[ \left( n^{(O)}_{\mathbf{q}} + 1 \right) \left( \eps - \hbar \, \omega^{(O)}_{\mathbf{q}} \right) H \left( \eps - \hbar \, \omega^{(O)}_{\mathbf{q}} -\eps_0 \right) \right. \\
	& \left. + n^{(O)}_{\mathbf{q}} \, \left( \eps + \hbar \, \omega^{(O)}_{\mathbf{q}} \right) H \left( \eps + \hbar \, \omega^{(O)}_{\mathbf{q}} -\eps_0 \right) \right] \\
	& = \frac{D_O^2}{\hbar^2 v_F^2 \sigma_m \omega_O} \left[ \left( n^{(O)}_{\mathbf{q}} + 1 \right) \left( \eps - \hbar \, \omega^{(O)}_{\mathbf{q}} \right) H \left( \eps - \hbar \, \omega^{(O)}_{\mathbf{q}} -\eps_0 \right) \right. \\
	& \left. + n^{(O)}_{\mathbf{q}} \, \left( \eps + \hbar \, \omega^{(O)}_{\mathbf{q}} \right) H \left( \eps + \hbar \, \omega^{(O)}_{\mathbf{q}} -\eps_0 \right) \right].
\end{align*}
Moreover, we can split the previous expression by considering the emission ($-$) and the absorption ($+$) cases separately, obtaining
\begin{align*}
	\Gamma^{(O^-)} & = \frac{D_O^2}{\hbar^2 v_F^2 \sigma_m \omega_O}\left(n^{(O)}_{\mathbf{q}}+1\right) \, \left( \eps - \hbar \, \omega^{(O)}_{\mathbf{q}} \right) H \left( \eps - \hbar \, \omega^{(O)}_{\mathbf{q}} -\eps_0 \right),\\
	\Gamma^{(O^+)} & = \frac{D_O^2}{\hbar^2 v_F^2 \sigma_m \omega_O}n^{(O)}_{\mathbf{q}} \, \left( \eps + \hbar \, \omega^{(O)}_{\mathbf{q}} \right) H \left( \eps + \hbar \, \omega^{(O)}_{\mathbf{q}} -\eps_0 \right).
\end{align*}

Similarly, if $\nu=K$ we have
\begin{align*}
	\Gamma^{(K)} & = \frac{1}{(\hbar v_F)^2} \frac{1}{(2\pi)^2} \frac{2\pi D_K^2}{\sigma_m \omega_O} 2\pi \left[ \left( n^{(K)}_{\mathbf{q}} + 1 \right) \left( \eps - \hbar \, \omega^{(K)}_{\mathbf{q}} \right) H \left( \eps - \hbar \, \omega^{(K)}_{\mathbf{q}} -\eps_0 \right) \right. \\
	& \left. + n^{(K)}_{\mathbf{q}} \, \left( \eps + \hbar \, \omega^{(K)}_{\mathbf{q}} \right) H \left( \eps + \hbar \, \omega^{(K)}_{\mathbf{q}} -\eps_0 \right) \right] \\
	& = \frac{D_K^2}{\hbar^2 v_F^2 \sigma_m \omega_K} \left[ \left( n^{(K)}_{\mathbf{q}} + 1 \right) \left( \eps - \hbar \, \omega^{(K)}_{\mathbf{q}} \right) H \left( \eps - \hbar \, \omega^{(K)}_{\mathbf{q}} -\eps_0 \right) \right. \\
	& \left. + n^{(K)}_{\mathbf{q}} \, \left( \eps + \hbar \, \omega^{(K)}_{\mathbf{q}} \right) H \left( \eps + \hbar \, \omega^{(K)}_{\mathbf{q}} -\eps_0 \right) \right].
\end{align*}
In the same way, we can split the previous expression by considering the emission ($-$) and the absorption ($+$) cases separately, obtaining
\begin{align*}
	\Gamma^{(K^-)} & = \frac{D_K^2}{\hbar^2 v_F^2 \sigma_m \omega_K}\left(n^{(K)}_{\mathbf{q}}+1\right) \, \left( \eps - \hbar \, \omega^{(K)}_{\mathbf{q}} \right) H \left( \eps - \hbar \, \omega^{(K)}_{\mathbf{q}} -\eps_0 \right),\\
	\Gamma^{(K^+)} & = \frac{D_K^2}{\hbar^2 v_F^2 \sigma_m \omega_K}n^{(K)}_{\mathbf{q}} \, \left( \eps + \hbar \, \omega^{(K)}_{\mathbf{q}} \right) H \left( \eps + \hbar \, \omega^{(K)}_{\mathbf{q}} -\eps_0 \right).
\end{align*}

At last, if $\nu=el-edg$ with the same change of variables introduced above we get
\begin{align*}
	\Gamma^{(el-edg)} & = \int_{\mathbb{R}^2} S^{(el-edg)}(\bk,\bk') \,d\bk' \\
	& = \frac{1}{(\hbar v_F)^2} \frac{N_i V_0^2}{2\pi\hbar W} \int_0^{2\pi} \int_{\eps_0}^{+\infty} \exp\left( -2\left(\sqrt{\left(\frac{\eps'}{\hbar v_F}\right)^2-\left(\frac{\pi}{W}\right)^2}\cos\theta'\right.\right.\\
	&\left.\left.-\sqrt{\left(\frac{\eps}{\hbar v_F}\right)^2-\left(\frac{\pi}{W}\right)^2}\cos\theta \right)^2 a^2 \right)\cdot \delta(\eps'-\eps)\eps' \, d\theta' \, d\eps'\\
	& = \frac{1}{(\hbar v_F)^2} \frac{N_i V_0^2}{2\pi\hbar W} \int_0^{2\pi} \int_{0}^{+\infty} \exp\left( -2\left(\sqrt{\left(\frac{\eps'}{\hbar v_F}\right)^2-\left(\frac{\pi}{W}\right)^2}\cos\theta'\right.\right.\\
	&\left.\left.-\sqrt{\left(\frac{\eps}{\hbar v_F}\right)^2-\left(\frac{\pi}{W}\right)^2}\cos\theta \right)^2 a^2 \right)\cdot \delta(\eps'-\eps)\eps' \chi(\eps'\in[\eps_0,+\infty[) \, d\theta' \, d\eps' \\
	& = \frac{1}{(\hbar v_F)^2} \frac{N_i V_0^2}{2\pi\hbar W} \int_0^{2\pi} \exp\left( -2\left(\sqrt{\left(\frac{\eps}{\hbar v_F}\right)^2-\left(\frac{\pi}{W}\right)^2}(\cos\theta'-\cos\theta) \right)^2 a^2 \right)\\
	& \cdot \eps H(\eps-\eps_0) \, d\theta' \\
	& = \frac{1}{(\hbar v_F)^2} \frac{N_i V_0^2}{2\pi\hbar W} \int_0^{2\pi} \exp\left( -2\left(\left(\frac{\eps}{\hbar v_F}\right)^2-\left(\frac{\pi}{W}\right)^2\right)(\cos\theta'-\cos\theta)^2 a^2 \right)\\
	& \cdot \eps H(\eps-\eps_0) \, d\theta'.
\end{align*}
The last integral can be easily computed numerically by the trapezoidal rule.

\bibliographystyle{abbrv}
\bibliography{NanoribbonsUQ.bib}

\end{document}